%% file: main.tex
\pdfoutput=1
\documentclass[11pt]{article}

\usepackage[preprint]{acl} 

\usepackage{times}
\usepackage{latexsym}
\usepackage[T1]{fontenc}
\usepackage[utf8]{inputenc}
\usepackage{microtype}
\usepackage{inconsolata}
\usepackage{graphicx}
\usepackage{lipsum}
\usepackage{geometry}
\usepackage{adjustbox}
\usepackage{booktabs}
\usepackage{float}
\usepackage{enumitem}
\usepackage{color}
\usepackage{stfloats}
\usepackage{placeins}
\usepackage{amsmath}
\usepackage{amsfonts}
\usepackage{kotex}
\usepackage{multirow}
\usepackage{array}
\usepackage{arydshln}
\usepackage{makecell}

\title{HEISIR: Hierarchical Expansion of Inverted Semantic Indexing for Training-free Retrieval of Conversational Data using LLMs}

\author{
    \textbf{Sangyeop Kim\textsuperscript{$\dagger$1,2}}, 
    \textbf{Hangyeul Lee\textsuperscript{2}}, 
    \textbf{Yohan Lee\textsuperscript{1}} \\
    \textsuperscript{1} Coxwave \\
    \textsuperscript{2} Seoul National University \\
    \small\texttt{sangyeop.kim@coxwave.com, mikelee@snu.ac.kr, yohan.lee@coxwave.com}
}

\begin{document}
\input{macros}

\maketitle
\renewcommand{\thefootnote}{$\dagger$}
\footnotetext{Corresponding author.}
\renewcommand{\thefootnote}{\arabic{footnote}}

\begin{abstract}
The growth of conversational AI services has increased demand for effective information retrieval from dialogue data. However, existing methods often face challenges in capturing semantic intent or require extensive labeling and fine-tuning. This paper introduces HEISIR (Hierarchical Expansion of Inverted Semantic Indexing for Retrieval), a novel framework that enhances semantic understanding in conversational data retrieval through optimized data ingestion, eliminating the need for resource-intensive labeling or model adaptation.
HEISIR implements a two-step process: (1) Hierarchical Triplets Formulation and (2) Adjunct Augmentation, creating semantic indices consisting of Subject-Verb-Object-Adjunct (SVOA) quadruplets. This structured representation effectively captures the underlying semantic information from dialogue content. HEISIR achieves high retrieval performance while maintaining low latency during the actual retrieval process. Our experimental results demonstrate that HEISIR outperforms fine-tuned models across various embedding types and language models. Beyond improving retrieval capabilities, HEISIR also offers opportunities for intent and topic analysis in conversational data, providing a versatile solution for dialogue systems.
\end{abstract}

\section{Introduction}

Conversational AI is being deployed across diverse industries, including personalized services \cite{kocaballi2019personalization}, code generation \cite{li2022competition}, educational tutoring \cite{mousavinasab2021intelligent} and even in social welfare services \cite{jo2023understanding}. This rapid expansion has resulted in vast amounts of dialogue data, rich with insights into user needs and behavioral patterns. To harness this wealth of information, Information Retrieval approach tailored to the unique characteristics of conversational data is necessary. We define this area as \textit{Conversational Data Retrieval} (CDR), which aims to enable efficient access to and extraction of relevant information from large-scale conversational data.

CDR is highly important for both end-users and conversational AI service providers. End-users frequently need to retrieve specific information from past conversations, such as previous agreements or discussion outcomes, using natural language queries. Service providers can utilize CDR to analyze user interactions and enhance their AI-powered services. For example, CDR can enhance Retrieval Augmented Generation (RAG) systems by providing relevant dialogue examples \cite{lewis2020retrieval, wang2024searching}, and support service improvement by identifying patterns in user behavior and recurring topics \cite{motger2022software, owoicho2022trec}.

However, conventional sparse and dense retrieval methods exhibit limited performance when applied to conversational data. Unlike traditional document data, conversational data has distinct characteristics that make traditional retrieval approaches insufficient. First, each utterance is assigned to a specific speaker, meaning that the same message can be expressed in entirely different ways depending on who is speaking. Additionally, a speaker's intention can be multifaceted; a single utterance may have multiple, often complex, intents in it. Furthermore, the meaning of utterances depends heavily on the context of the dialogue and the relationship between participants, often without explicit contextual markers. These unique characteristics pose significant challenges that existing context-aware retrieval models, trained primarily on document-based QA tasks \cite{yang2015wikiqa, rajpurkar2016squad, nguyen2016ms}, struggle to address.

\begin{figure*}[!htbp]
    \centering
    \includegraphics[width=1.0\textwidth]{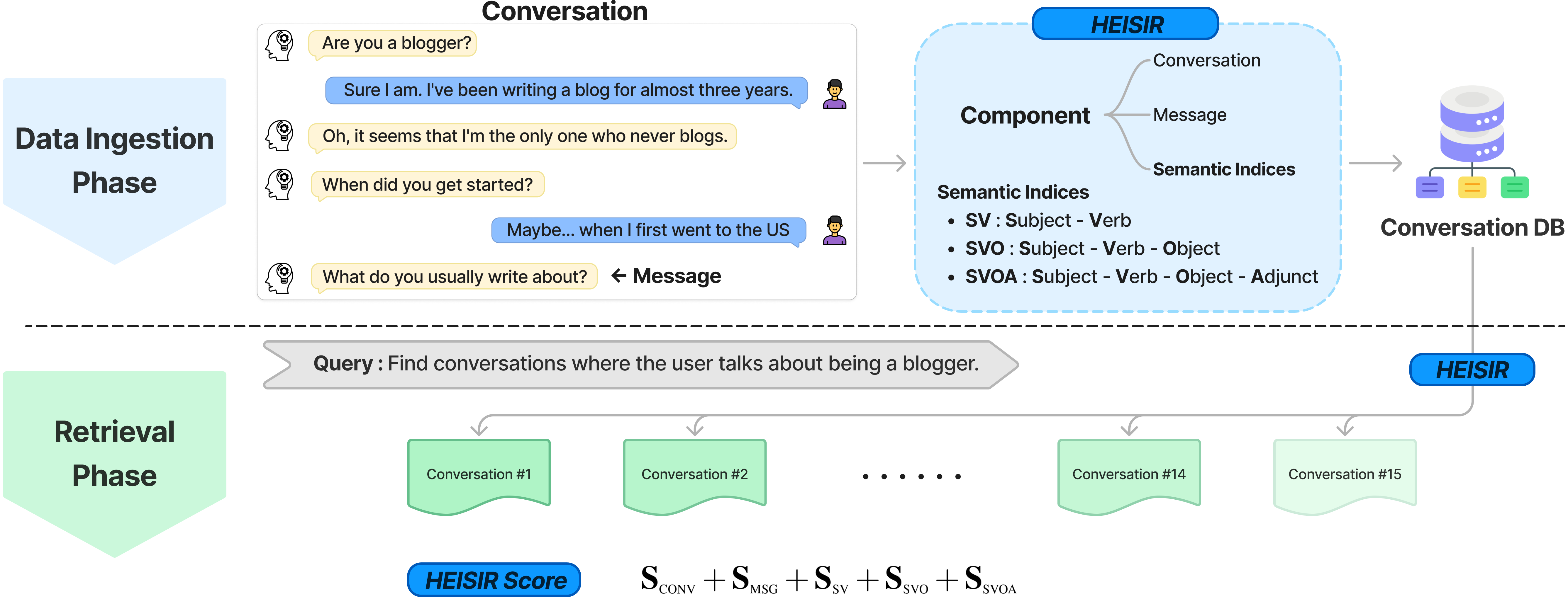}
    \caption{Architecture of HEISIR framework: Data Ingestion Phase}
    \label{fig:phase}
\end{figure*}

Apart from the limited effectiveness of existing retrieval methods on conversational data, the structure of the retrieval system itself poses a significant challenge to practical implementation. A typical retrieval system consists of two main phases \cite{wang2021milvus}: the data ingestion phase, where incoming data is processed and indexed offline, and the retrieval phase, where relevant information is searched based on user queries. Recent research on retrieval system leverages the powerful context understanding capability of LLMs through techniques such as re-ranking \cite{zhang2023rankinggpt}, query rewriting \cite{yu2020few}, and using larger retrieval models \cite{repllama, peng2023soft}. However, these approaches, especially when incorporating LLMs, introduce significant latency trade-off in the retrieval phase, making them impractical for real-time services.

Building on these insights, we propose a novel framework that extracts and processes the inverted semantic indices in the data ingestion phase, unlike traditional approaches that process indices in the retrieval phase. \textbf{HEISIR} (\textbf{H}ierarchical \textbf{E}xpansion of \textbf{I}nverted \textbf{S}emantic \textbf{I}ndexing for \textbf{R}etrieval) implements structured semantic indices that capture the inherent syntactic hierarchy within sentences. Overall scheme of our framework is detailed in Figure \ref{fig:phase}. HEISIR constructs search indices based on the inherent syntax present in all natural language sentences, eliminating the need for extensive labeling and training. In data ingestion phase, HEISIR extracts semantic indices and stores them as \textit{inverted indices}. In retrieval phase, HEISIR computes score to retrieve the most relevant conversations. Our approach not only enhances retrieval performance but also offers practical advantages, as it eliminates latency during the retrieval phase. The key contributions of this research are:

\begin{center}
\begin{minipage}{0.95\linewidth}
\begin{enumerate}[itemsep=0pt, topsep=0pt, parsep=5pt, partopsep=0pt, leftmargin=10pt]
\item \textbf{Improved retrieval performance} significantly enhances retrieval capabilities with only negligible increase of latency by optimizing data ingestion.
\item \textbf{Practical real-world applicability} enables deployment of effective dialogue retrieval systems in production environments lacking labeled training data.
\item \textbf{Practical-scale Robustness} consistently enhances performance when integrated with any combination of language model scales.
\item \textbf{Versatility beyond retrieval} provides highly interpretable atomic semantic units, enabling intuitive intent and topic analysis.
\end{enumerate}
\end{minipage}
\end{center}

\section{Linguistic Preliminaries}
\label{section:linguistic}

To design semantic indices for CDR, it is necessary to identify two fundamental components in a message: the speaker and the intent. These components correspond to the subject and the verb of a sentence. To capture complex intents while maintaining the structural integrity, HEISIR follows the syntactic comprehension process of humans.

\begin{figure}[!htbp]
    \centering
    \includegraphics[width=0.90\linewidth]{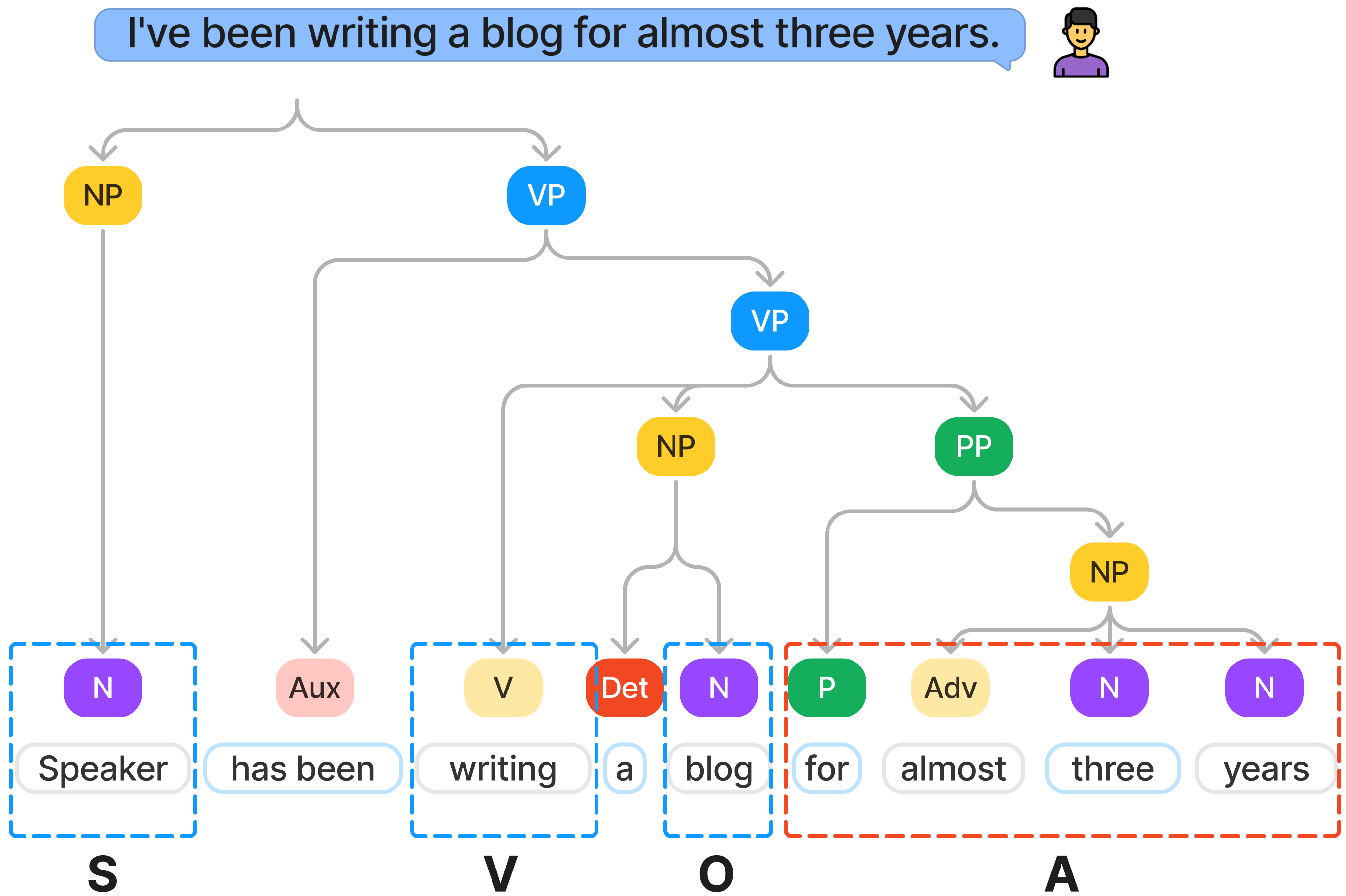}
    \caption{Phrase Structure Grammar and Constituents}
    \label{fig:psg}
\end{figure}

Numerous studies support the view that humans comprehend sentences incrementally. Specifically, sentences are processed by first constructing the simplest syntactic structure and then integrating semantic adjuncts to achieve full understanding \cite{kamide2003time, altmann2009incrementality, fossum2012sequential}. \textbf{Phrase Structure Grammar (PSG)} \cite{chomsky2002syntactic, gazdar1985generalized} is one of the most widely used models to explain this top-down syntactic processing. PSG breaks down natural language sentences into \textbf{constituents}—syntactically significant units—organized in a hierarchical binary tree structure. The overall scheme of how PSG parses messages for HEISIR is detailed in Figure \ref{fig:psg}. 

Another key concept central to syntactic processing is \textbf{verb valency}. Verb valency refers to the number of arguments that a verb can take in a sentence, and it describes the relationship between the verb and other elements in the clause, such as the subject, direct object, and indirect object. Verbs can be categorized based on their valency as \textbf{avalent} (taking no arguments), \textbf{monovalent} (taking one argument), \textbf{divalent} (taking two arguments), or \textbf{trivalent} (taking three arguments). Verb valency plays a critical role in determining the constituent hierarchy in sentences, as different valency types lead to different phrase structures.

In this study, we develop a semantic indexing framework according to hierarchical syntactic processing schemes that represent the three most common types of verb valency: monovalent, divalent, and trivalent. We parse the constituents from sentences as follows:
\begin{itemize}
    \item \textbf{Subject}: The entity performing the action or the one that the sentence is about.
    \item \textbf{Verb}: The action or state described in the sentence.
    \item \textbf{Object}: The entity directly affected by the verb’s action.
    \item \textbf{Adjunct}: An optional or necessary element that provides additional information about the action, such as time, place, or manner.
\end{itemize}

HEISIR borrows these concepts from syntax, but uses them in a slightly different context. For instance, HEISIR fixes the \textbf{Subject} to the speaker of utterance, to perform information retrieval in conversational data. Therefore, avalent verbs are disregarded due to the existence of an explicit subject. This makes index tuples include at least two constituents: the \textbf{Subject} and the \textbf{Verb}. Furthermore, we use the word \textbf{Adjunct} in a slightly broader context than is conventional; For divalent verbs that take two objects, it is syntactically correct to allocate both direct and indirect objects under the \textbf{Object} constituent. Instead, we include indirect object into the \textbf{Adjunct} category for structural integrity and better search performance.

\section{Hierarchical Expansion of Inverted Semantic Indexing for Retrieval}

\begin{figure}[!htbp]
    \centering
    \includegraphics[width=0.95\linewidth]{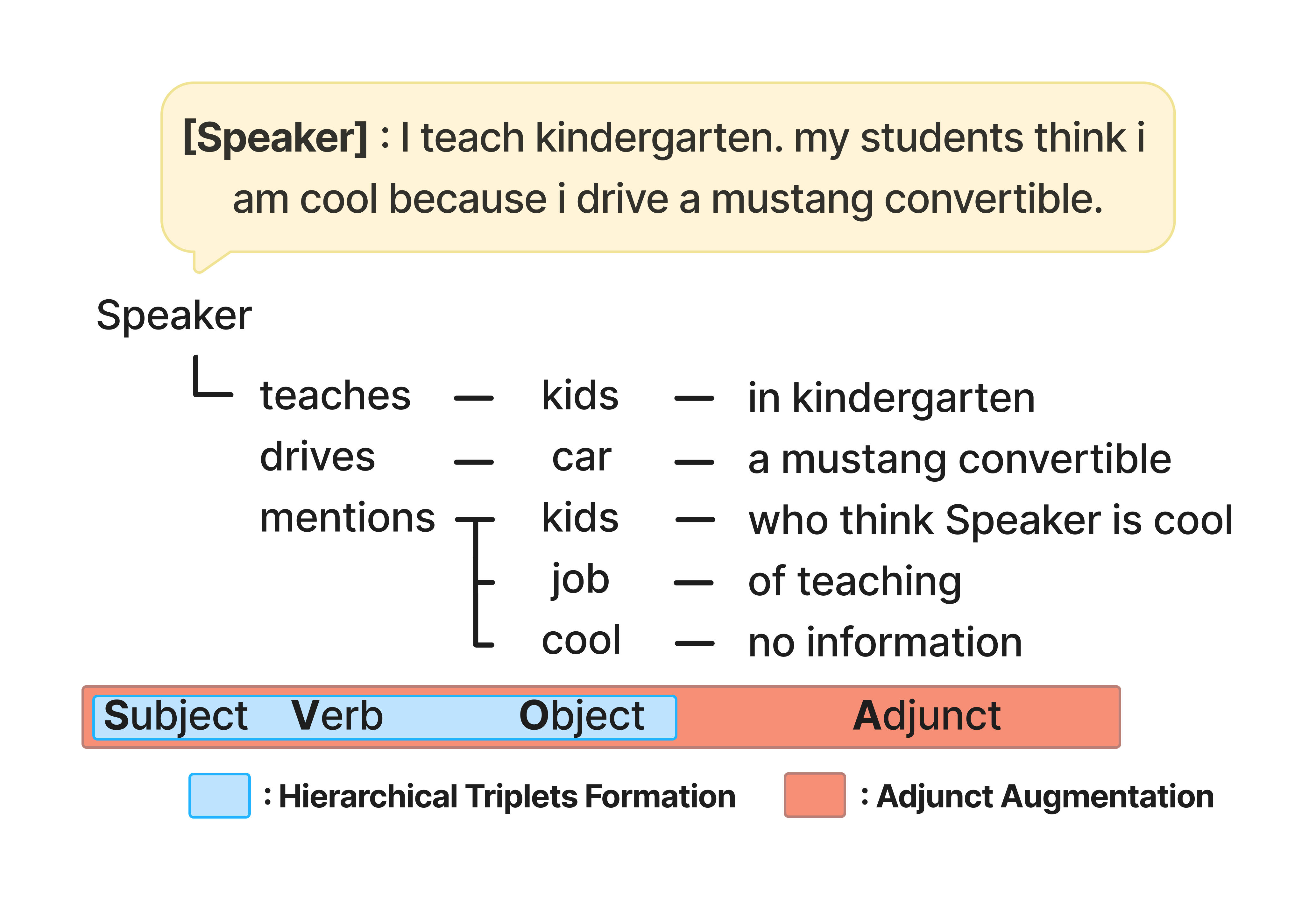}
    \caption{2-Step Expansion Process of HEISIR}
    \label{fig:triplet_detail}
\end{figure}

HEISIR is a novel conversational data retrieval framework that incorporates incremental syntactic processing with inverted indexing. The framework invests resources in the data ingestion phase, allowing for optimized performance during the retrieval phase without sacrificing latency.

Figure \ref{fig:triplet_detail} outlines two key steps of the HEISIR framework: (1) \textbf{Hierarchical Triplets Formulation}, identifying the fundamental syntactic constituents -- subject, verb, and object -- forming the core structures of the sentence; (2) \textbf{Adjunct Expansion}, expanding SVO triplets to SVOA quadruplets, where A represents an Adjunct. 

While a single-step approach for SVOA quadruplet extraction may seem straightforward, it has critical limitations in accuracy and control. For example, it generates redundant variations like ``teaches kids at kindergarten'' and ``teaches children in kindergarten'' that carry the same meaning, introducing unnecessary index noise.

In contrast, HEISIR's two-step approach first extracts core SVO triplets and then carefully adds adjuncts, effectively preventing redundancy while maintaining high-quality indices. Furthermore, the experimental results supporting this analysis are presented in Section \ref{subsubsec-2-step-index-construction}. The detailed prompts used in our approach can be found in Appendix \ref{apdx:prompts}.

\subsection{Data Ingestion Phase}
\paragraph{Step 1: Hierarchical Triplets Formulation} 
In this step, HEISIR breaks down sentences into one or more \textbf{SVO triplets}. These triplets capture the core syntactic hierarchy within the dialogue content, addressing two key limitations of traditional embedding methods \cite{word2vec, glove}. First, HEISIR reduces ambiguity in the embeddings of complex sentences. HEISIR decomposes messages into multiple SVO triplets, significantly reducing sentence complexity. Additionally, HEISIR strictly excludes pronouns to ensure semantic completeness in each index. Second, our approach minimizes the impact of non-semantic elements on performance. As discussed in section \ref{section:linguistic}, HEISIR fixes the subject to the speaker of the message to enhance retrieval performance. During this syntactic transformation, avalent verbs are transformed into their mono-, di-, or trivalent equivalents, and all tenses, auxiliaries, and syntactic markers are removed. This process refines HEISIR-generated triplets to capture only semantically significant elements of the message.

\paragraph{Step 2: Adjunct Augmentation}
SVO triplet indices extracted in step 1 identify hierarchical structures in the message. To enhance specificity, we introduce a \textit{Adjunct Augmentation} process.
We categorize four distinct patterns of adjuncts:

\begin{enumerate}[label={-}, leftmargin=*, itemsep=0pt, topsep=5pt, parsep=2pt, partopsep=0pt]
\item \textbf{Detailed Content and Theme of Discussion} specifies the target of communication or conversational themes using prepositions.
\item \textbf{Reason or Causation} explains the underlying causes or reasons for actions or situations, addressing the "why" of a scenario.
\item \textbf{Condition or Accompanying Circumstance} describes the context or conditions under which actions occur or situations exist.
\item \textbf{No information} indicates cases where adding detailed information is impossible or not meaningful for the given sentence structure.
\end{enumerate}

Following this step, each message in a conversation is encoded into one or more SVOA quadruplets. These SVOA quadruplets break down complex messages into comprehensible semantic units. The indices are then stored in an inverted index structure, which facilitates efficient retrieval in the subsequent phase.

\subsection{Retrieval Phase: Scoring}

After deriving SVOA quadruplets from a conversation, we devise a method to collectively evaluate embeddings of five conversational components --- conversation, message, SV, SVO, SVOA --- in a single score metric.

To formalize our scoring method, we represent queries and conversations with embeddings \( E_q \) and \( E_{\text{conv}} \) respectively, while other conversational components \(C=\{ \text{message, SV, SVO, SVOA} \}\) are encoded as \( E_c \). The relevance between components is computed using a vector similarity function \( f(\cdot,\cdot) \). The scoring process is defined as follows:

\begin{equation}
\label{eq:s_conv}
S_{\text{conv}} = f(E_q, E_{\text{conv}})
\end{equation}
\\[-0.6cm]
\begin{equation}
\label{eq:s_c}
\begin{gathered}
S_{c} = \max_{E_{c_m}} f(E_q, E_{c_m}) \text{ for } c \in C\\
\text{where } E_{c_m} \in \mathbb{E}_{c}(m), m \in \text{conv}
\end{gathered}
\end{equation}
\\[-0.4cm]
\begin{equation}
\label{eq:s_heisir}
S_{\text{HEISIR}} = S_{\text{conv}} + \sum_{c \in C} S_c
\end{equation}

Equation (\ref{eq:s_conv}) measures the semantic similarity between the query and the entire conversation content to capture conversation-level relevance. Component scores are computed as shown in (\ref{eq:s_c}), where we evaluate each type separately by finding the highest similarity between the query and component instances within the conversation. For each conversational component \(c \in C\), \(\mathbb{E}_{c_m}\) denotes a set of embeddings of component \(c\) in the message \(m\), and \(\text{max}_{E_{c_m}}\) picks best from multiple <component, message> pairs. Finally, the overall HEISIR score in (\ref{eq:s_heisir}) is determined by aggregating the conversation-level similarity with individual component scores to provide a comprehensive measure of relevance.

Potential alternatives of maximization function in (\ref{eq:s_c}) are summation and averaging. However, we employ maximization instead of these alternatives since summation and averaging introduce noise from less relevant messages, whereas focusing on the most salient component is intuitively more effective.

\section{Experiment}
\subsection{Dataset}
We select five dialogue datasets addressing key topics in conversational data analysis: profanity detection, user satisfaction evaluation, and personal information protection. We extract queries from five datasets and map them to corresponding conversation sessions. The statistics of the derived dataset is detailed in Table \ref{tab:dataset-stats}.

\begin{enumerate}[label={-}, leftmargin=*, itemsep=3pt, topsep=5pt, parsep=2pt, partopsep=0pt]
\item \textbf{BAD} \cite{databad} focuses on improving safety of conversational agents.
\item \textbf{DICES} \cite{datadices} evaluates safety of AI responses in diverse contexts.
\item \textbf{Daily Dialog} \cite{datadaily} provides labeled multi-turn dialogues reflecting user sentiments. 
\item \textbf{PILD} \cite{datapild} detects and protects sensitive personal information.
\item \textbf{USS} \cite{datauss} simulates user satisfaction in task-oriented dialogue systems.
\end{enumerate}

\newcommand{\betweensmallscript}{\fontsize{8.0}{10.2}\selectfont}
\begin{table}[htbp]
\centering
\betweensmallscript
\renewcommand{\arraystretch}{1.2}
\begin{tabular}{@{}c@{\hspace{1em}}c@{\hspace{1em}}c@{\hspace{1em}}c@{\hspace{1em}}c@{}}
\toprule
\textbf{Dataset} & \textbf{Conv.} & \textbf{Query} & \makecell{\textbf{Utterances}\\\textbf{per Conv.}} & \makecell{\textbf{Mapped Conv.}\\\textbf{per Query}} \\
\midrule
Train & \multirow{2}{*}{4,096} & 3,000 & \multirow{2}{*}{12.3} & 15.5 \\
Validation &  & 590 &  & 14.9 \\
\addlinespace[0.3em]
\hline
\addlinespace[0.3em]
Test & 4,035 & 3,501 & 12.7 & 15.4 \\
\bottomrule
\end{tabular}
\caption{Statistics of Dataset}
\label{tab:dataset-stats}
\end{table}

\subsection{Baseline}
For the evaluation of Information Retrieval, we compare the following baseline models:
\begin{enumerate}[label={-}, leftmargin=*, itemsep=3pt, topsep=5pt, parsep=2pt, partopsep=0pt]
\item {\textbf{DPR}} \cite{dpr}: Learns dense embeddings for queries and passages using a dual-encoder architecture for efficient retrieval.
\item {\textbf{SPLADE-v3}} \cite{lassance2024splade}: Employs sparse lexical representations for expansion-based retrieval, enhancing earlier SPLADE versions \cite{formal2021splade}.
\item \textbf{LLM2Vec} \cite{embedllm2vec}: Creates dense vector representations of text using large language models to facilitate retrieval tasks.
\end{enumerate}

For training-free models, we utilize the following in our comparison:
\begin{enumerate}[label={-}, leftmargin=*, itemsep=3pt, topsep=5pt, parsep=2pt, partopsep=0pt]
\item \textbf{CoT Expansion} \cite{jagerman2023query}: Expands queries using CoT prompting and uses both original and expanded queries for retrieval.
\item \textbf{HyDE} \cite{gao2022precise}: Generates hypothetical relevant documents using an LLM to enhance retrieval performance.
\item \textbf{LameR} \cite{shen2023large}: Produces hypothetical documents using BM25 initial results to improve retrieval effectiveness.
\end{enumerate}

\subsection{Experimental Setup}

For context consideration, we set a window of \(k=2\) previous messages. We set LLMs temperature to 0.0, and use cosine similarity for embedding comparisons. All implementation details, including the versions of the embedding models and LLMs used, can be found in Appendix \ref{apdx:reproducing}.

\paragraph{Embedding for HEISIR}
For this study, We select embedding models based on their performance in the MTEB benchmark \cite{mtebleaderboard}. We explore Encoder-only models with extended context capabilities, such as GTE \cite{embedGTE} and Nomic \cite{embednomic}. Another category of embedding models is decoder-only models based on LLMs. Specifically, we utilize LLM2Vec \cite{embedllm2vec} with its LLama-3 \cite{llmllama3} based version. Additionally, we use NV-Embed \cite{embednv}, which currently achieves the highest average MTEB benchmark score. For comprehensive evaluation, We employ latest OpenAI-small and large embedding models \cite{embedopenai}.

\paragraph{LLMs for HEISIR}
The study uses a range of LLM models for different computational environments. For low-resource settings, we assess Gemma (2B) \cite{llmgemma}, Phi-3 (mini, 3.8B) \cite{llmphi3}, LLama-3 (8B) \cite{llmllama3}, and Qwen-2 (7B) \cite{llmqwen2}, which are suitable for environments with limited hardware. We also test API-based models including GPT-3.5-turbo \cite{llmgpt35} and Claude 3 Haiku \cite{llmhaiku}. 

\section{Result}

\subsection{Baseline Result}

\begin{table}[!htbp]
\centering

\resizebox{\columnwidth}{!}{%
\footnotesize
\begin{tabular}{@{}llcccc@{}}
\toprule
\textbf{Type} & \textbf{Model} & \textbf{acc@1} & \textbf{ndcg@5} & \textbf{ndcg@10} & \textbf{ndcg@20} \\
\midrule
\multirow{3}{*}{Baseline} 
& DPR         & 0.0337 & 0.0274 & 0.0257 & 0.0283 \\
& SPLADE-v3   & 0.2048 & 0.1696 & 0.1614 & 0.1743 \\
& LLM2Vec     & 0.2825 & 0.2225 & 0.2092 & 0.2220 \\
\midrule
\multirow{3}{*}{Fine-tuned} 
& DPR         & 0.1708 & 0.1420 & 0.1342 & 0.1444 \\
& SPLADE-v3   & 0.2474 & 0.2015 & 0.1908 & 0.2034 \\
& LLM2Vec     & 0.3533 & 0.2881 & 0.2745 & 0.2912 \\
\midrule
\multirow{2}{*}{\makecell[c]{HEISIR}} & \makecell[c]{GPT-3.5-turbo} & \multirow{2}{*}{\makecell[c]{\textbf{0.4085}}} & \multirow{2}{*}{\makecell[c]{\textbf{0.3260}}} & \multirow{2}{*}{\makecell[c]{\textbf{0.3056}}} & \multirow{2}{*}{\makecell[c]{\textbf{0.3198}}} \\
& \makecell[c]{+ OpenAI-large} & & & & \\
\bottomrule
\end{tabular}%
}
\caption{Performance metrics of baseline, fine-tuned, and best models}
\label{tab:model-performance-comparison}
\end{table}

\begin{table*}[t]
\centering
\small
\setlength{\tabcolsep}{4pt}
\resizebox{0.99\textwidth}{!}{%
\begin{tabular}{ccc*{12}{r}}
\toprule
\multicolumn{3}{c}{\multirow{2}{*}{\makecell{\textbf{Model}}}} & \multicolumn{4}{c}{GTE (Encoder-only)} & \multicolumn{4}{c}{LLM2Vec (Decoder-only)} & \multicolumn{4}{c}{OpenAI-large (API)} \\
\cmidrule(lr){4-7} \cmidrule(lr){8-11} \cmidrule(lr){12-15}
& & & acc@1 & ndcg@5 & ndcg@10 & ndcg@20 & acc@1 & ndcg@5 & ndcg@10 & ndcg@20 & acc@1 & ndcg@5 & ndcg@10 & ndcg@20 \\
\midrule
\multirow{2}{*}{\rotatebox[origin=c]{90}{Base}} & Pre-trained & & 0.3156 & 0.2550 & 0.2394 & 0.2520 & 0.2825 & 0.2225 & 0.2092 & 0.2220 & 0.3425 & 0.2826 & 0.2670 & 0.2839 \\
 & Fine-tuned & & 0.3708 & 0.2994 & 0.2823 & \textbf{0.2984} & 0.3533 & 0.2881 & 0.2745 & 0.2912 & - & - & - & - \\
\midrule
\multirow{2}{*}{\rotatebox[origin=c]{90}{Mini}} & Gemma & & 0.3096 & 0.2562 & 0.2415 & 0.2541 & 0.3136 & 0.2597 & 0.2428 & 0.2543 & 0.3422 & 0.2733 & 0.2574 & 0.2705 \\
& Phi-3 & & 0.3582 & 0.2934 & 0.2757 & 0.2901 & \underline{0.3865} & \underline{0.3138} & \underline{0.2936} & \underline{0.3055} & 0.3950 & 0.3192 & 0.2992 & 0.3136 \\
\midrule
\multirow{2}{*}{\rotatebox[origin=c]{90}{Small}} & LLama-3 & & \underline{\textbf{0.3728}} & \underline{\textbf{0.3005}} & \underline{\textbf{0.2829}} & 0.2958 & \underline{0.3902} & \underline{0.3155} & \underline{0.2944} & \underline{0.3073} & 0.4053 & \textbf{0.3263} & \textbf{0.3072} & \textbf{0.3207} \\
 & Qwen-2 & & 0.3613 & 0.2952 & 0.2786 & 0.2921 & \underline{0.3879} & \underline{0.3123} & \underline{0.2912} & \underline{0.3035} & 0.4045 & 0.3224 & 0.3026 & 0.3166 \\
\midrule
\multirow{2}{*}{\rotatebox[origin=c]{90}{API}} & GPT-3.5-turbo & & 0.3633 & 0.2943 & 0.2776 & 0.2930 & \underline{0.3916} & \underline{0.3140} & \underline{0.2934} & \underline{0.3068} & \textbf{0.4085} & 0.3260 & 0.3056 & 0.3198 \\
 & Haiku & & \underline{0.3716} & 0.2978 & 0.2809 & 0.2943 & \underline{\textbf{0.3927}} & \underline{\textbf{0.3161}} & \underline{\textbf{0.2958}} & \underline{\textbf{0.3086}} & 0.4056 & 0.3245 & 0.3028 & 0.3185 \\
\bottomrule
\end{tabular}%
}
\caption{Performance comparison of different models across Encoder-only (GTE), Decoder-only (LLM2Vec), and API (OpenAI-large) approaches: Underlined values outperform the Fine-tuned model, bold values indicate the best performing LLM model combination for each embedding model.}

\label{tab:main_result}
\end{table*}

As shown in Table \ref{tab:model-performance-comparison}, models that typically excel in general document retrieval struggle to achieve high performance in conversational data retrieval. The LLM2Vec model \cite{embedllm2vec},  which is based on a decoder-only architecture specialized for dialogue generation, outperforms other models, indicating that embedding-based retrieval methods are more effective for semantic search in the context of conversational data. Fine-tuning alone does not sufficiently address this issue, suggesting the need for retrieval methods specifically tailored to conversational data.

\subsection{Main Result}
In this section, we only discuss the best-performing combinations of LLMs and embedding models. Results for all other combinations also show the effectiveness of HEISIR, with details available in Appendix \ref{apdx:results}. 

Table \ref{tab:main_result} demonstrates the effectiveness of HEISIR method in conversational data retrieval across various model architectures and embedding types. Our model, in combination with various LLMs, outperforms all pre-trained encoder-only and decoder-only baselines and API models, except for when paired with Gemma. Notably, many of these models even surpass the fine-tuned baselines, highlighting the key strength of our method: achieving high performance without a need for resource-intensive data labeling and training. These results provide insights into optimal combinations for various constraints. OpenAI-large with GPT-3.5-turbo achieve the highest overall performance. For environments prohibiting external APIs, LLama-3 + LLM2Vec perform best, though other combinations show comparable results, offering flexibility. In resource-constrained settings, LLama-3 + GTE provides an efficient balance of performance and resource usage.

\subsection{Practical Retrieval Efficiency}

\label{subsection:practical}

\begin{table}[h]
\centering
\normalsize
\resizebox{\columnwidth}{!}{%
\begin{tabular}{l c c c c r}
\toprule
\textbf{Model} & \textbf{acc@1} & \textbf{ndcg@5} & \textbf{ndcg@10} & \textbf{ndcg@20} & \textbf{Time (s)} \\
\toprule
BM25 & 0.1465 & 0.1233 & 0.1200 & 0.1299 & 0.0045 \\
\midrule
CoT Expansion & 0.1225 & 0.0963 & 0.0894 & 0.0926 & 2.0059 \\
HyDE & 0.2702 & 0.2139 & 0.2026 & 0.2128 & 2.6192 \\
LameR & 0.2245 & 0.1853 & 0.1752 & 0.1866 & 3.5465 \\
\midrule
LLM2Vec & 0.2825 & 0.2225 & 0.2092 & 0.2220 & 0.2091 \\
$\text{HEISIR}_{\text{LLM2Vec}}$ & 0.3902 & 0.3155 & 0.2944 & 0.3073 & 0.2778 \\
OpenAI-large & 0.3425 & 0.2826 & 0.2670 & 0.2839 & 0.5966 \\
$\text{HEISIR}_{\text{OpenAI-large}}$ & 0.4085 & 0.3260 & 0.3056 & 0.3198 & 0.7334 \\
\bottomrule
\end{tabular}%
}
\caption{Performance Comparison of Different Models}
\label{tab:time_comparison}
\end{table}

Table \ref{tab:time_comparison} demonstrates that HEISIR outperforms various training-free methods in both speed and accuracy. Methods that rely on LLMs for query expansion during the search phase, such as CoT Expansion \cite{jagerman2023query}, LameR \cite{lamer}, and HyDE \cite{hyde}, require significantly longer search times. In contrast, HEISIR adds minimal time during retrieval phase (0.0687 seconds to LLM2Vec, 0.1367 seconds to OpenAI-large) by requiring only simple score computation. This efficiency is achieved by shifting the computational load to the data ingestion phase, which takes 3.86 seconds with GPT-3.5-turbo and 1.22 seconds with LLama-3. These ingestion times are reasonable considering they can be significantly reduced through parallelization techniques. Additionally, HEISIR is more cost-effective. While other methods incur costs per search, HEISIR only requires a one-time data processing step during the data ingestion phase. As search frequency increases, HEISIR remains economically efficient, whereas costs for other methods continue to rise. LLM inference for indexing incurs an additional cost of approximately \$0.00894 per conversation, based on the latest API pricing.

An interesting trend was observed in the experimental results: LameR, which is designed as an improvement over HyDE, shows inferior performance than HyDE. LameR integrates BM25 for higher retrieval performance. This unexpected result suggests that while BM25 may be well-suited for traditional document retrieval, it appears to hinder performance in conversational data retrieval.

\section{Analysis}
\subsection{Interacting with Each Component}
\label{subsection:component}

\begin{figure}[htbp]
    \centering
    \includegraphics[width=0.82\linewidth]{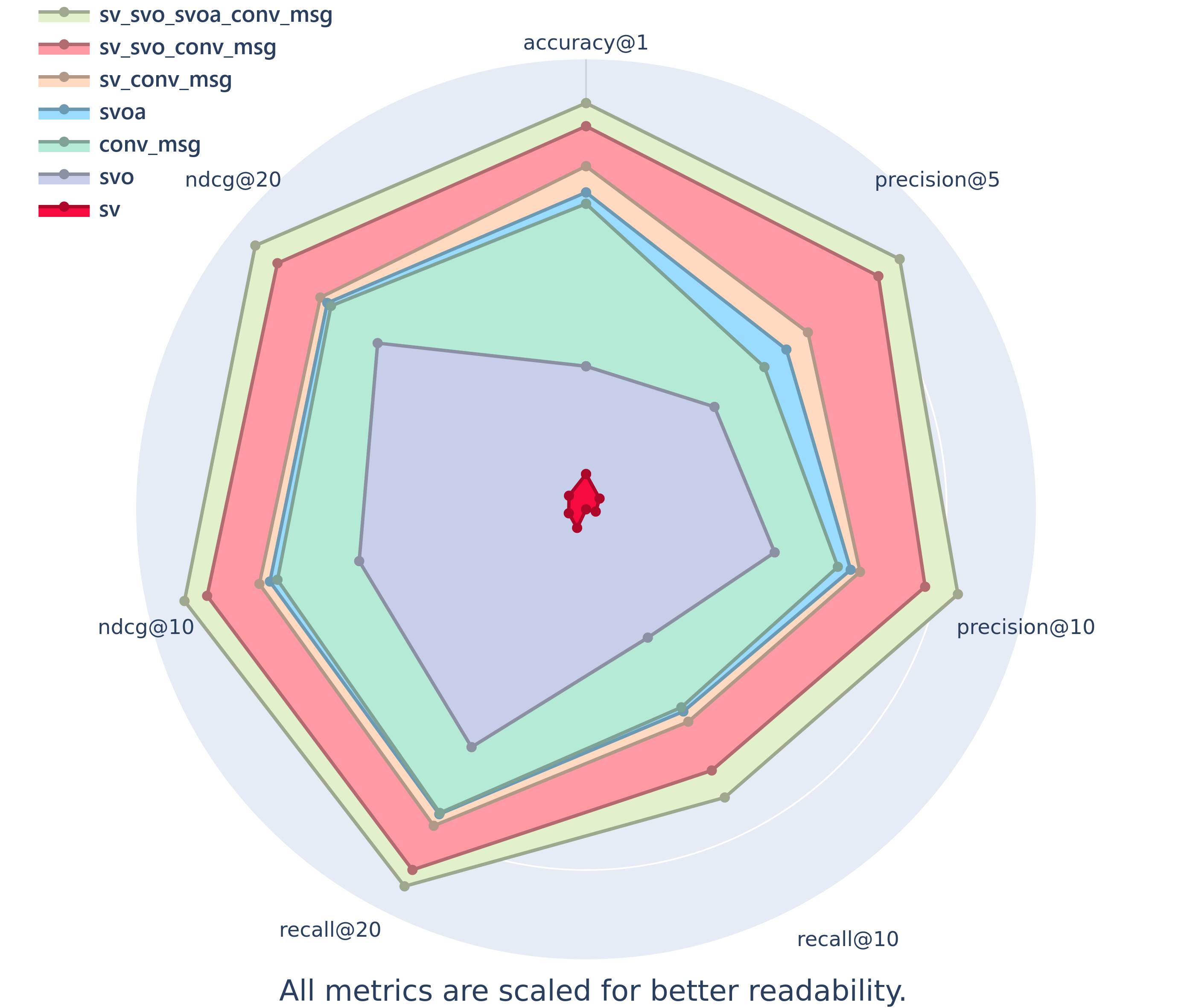}
    \caption{Marginal Performance of Components}
    \label{fig:enter-label}
\end{figure}

Figure \ref{fig:enter-label} shows the average performance across all experimental settings. Each contour represents a combination of the conversational components: Conversation, Message, SV, SVO, SVOA. 
Notably, SVOA index embeddings outperforms existing retrieval methods that rely on conversation and message embeddings. The best performance is achieved when embeddings of all conversational components are combined, demonstrating the effectiveness of HEISIR.

\subsection{2-step index construction} \label{subsubsec-2-step-index-construction}

\begin{table}[htbp] 
\centering
\resizebox{\columnwidth}{!}{%
\scriptsize
\setlength{\tabcolsep}{1.5pt}
\renewcommand{\arraystretch}{0.8}
\begin{tabular}{cccccc}
\toprule
\multicolumn{2}{c}{\textbf{Model}} & \multicolumn{4}{c}{\textbf{Metric}} \\
\cmidrule[0.8pt](r{0.125em}){1-2} \cmidrule[0.8pt](l{0.125em}){3-6}
\textbf{LLM} & \textbf{embedding} & \textbf{acc@1} & \textbf{ndcg@5} & \textbf{ndcg@10} & \textbf{ndcg@20} \\
\midrule
\multirow{9}{*}[+0.2em]{\rotatebox[origin=c]{90}{\textbf{Phi-3}}} 
& \multirow{2}{*}{\centering GTE} & 0.3530 & 0.2883 & 0.2724 & 0.2856 \\
& & (-1.45\%) & (-1.74\%) & (-1.20\%) & (-1.55\%) \\
\cmidrule{2-6}
& \multirow{2}{*}{\centering LLM2Vec} & 0.3405 & 0.2724 & 0.2559 & 0.2692 \\
& & \textbf{(-11.90\%)} & \textbf{(-13.19\%)} & \textbf{(-12.84\%)} & \textbf{(-11.88\%)} \\
\cmidrule{2-6}
& \multirow{2}{*}{\centering OpenAI-large} & 0.3879 & 0.3149 & 0.2970 & 0.3137 \\
& & (-1.80\%) & (-1.35\%) & (-0.74\%) & (+0.03\%) \\
\midrule
\multirow{9}{*}[+0.2em]{\rotatebox[origin=c]{90}{\textbf{LLama-3}}} 
& \multirow{2}{*}{\centering GTE} & 0.3513 & 0.2872 & 0.2713 & 0.2843 \\
& & \textbf{(-5.77\%)} & (-4.43\%) & (-4.10\%) & (-3.89\%) \\
\cmidrule{2-6}
& \multirow{2}{*}{\centering LLM2Vec} & 0.3362 & 0.2710 & 0.2544 & 0.2693 \\
& & \textbf{(-13.84\%)} & \textbf{(-14.11\%)} & \textbf{(-13.59\%)} & \textbf{(-12.37\%)} \\
\cmidrule{2-6}
& \multirow{2}{*}{\centering OpenAI-large} & 0.3865 & 0.3106 & 0.2957 & 0.3123 \\
& & (-4.64\%) & (-4.81\%) & (-3.74\%) & (-2.62\%) \\
\midrule
\multirow{9}{*}[+0.2em]{\rotatebox[origin=c]{90}{\textbf{GPT-3.5-turbo}}} 
& \multirow{2}{*}{\centering GTE} & 0.3510 & 0.2844 & 0.2682 & 0.2823 \\
& & (-3.39\%) & (-3.36\%) & (-3.39\%) & (-3.65\%) \\
\cmidrule{2-6}
& \multirow{2}{*}{\centering LLM2Vec} & 0.3359 & 0.2696 & 0.2533 & 0.2673 \\
& & \textbf{(-14.22\%)} & \textbf{(-14.14\%)} & \textbf{(-13.67\%)} & \textbf{(-12.88\%)} \\
\cmidrule{2-6}
& \multirow{2}{*}{\centering OpenAI-large} & 0.3876 & 0.3126 & 0.2954 & 0.3111 \\
& & \textbf{(-5.12\%)} & (-4.11\%) & (-3.34\%) & (-2.72\%) \\
\bottomrule
\end{tabular}%
}
\caption{Single-step Performance (\% Indicates Performance Change Compared to 2-step)}
\label{tab:1step_2step}
\end{table}

The key strength of HEISIR lies in its 2-step SVOA quadruplet extraction process. This approach offers both qualitative and quantitative advantages over single-step extraction methods. First, the 2-step approach extracts SVOA quadruplets with greater precision. The 
\textbf{Hierarchical Triplet Formulation} step establishes the syntactic hierarchy within the message, while the \textbf{Adjunct Augmentation} step collects detailed information, simulating the incremental sentence comprehension process of humans. This method consistently produces higher-quality SVOA quadruplets compared to single-step extraction. Second, quantitative evaluations show that our 2-step approach consistently outperforms the single-step method (Table \ref{tab:1step_2step}).

\subsection{Exploring the effects of score ensembles}
While our results confirm that combining the extracted semantic indices enhances retrieval performance, this improvement can be claimed as a simple \textit{ensemble effect}. In this subsection, we compare the ensemble effect with HEISIR. To measure the ensemble effect, we combine scores from all embedding models for the conversation and message components, excluding the significantly underperforming NV-Embed model \cite{embednv} to avoid underestimating the ensemble effect. For comparison, we apply the HEISIR method to the Phi-3 model \cite{llmphi3}, which is the smallest and least performant among our HEISIR variants. As shown in Table \ref{tab:ensemble}, HEISIR outperforms the ensemble effect, even in its least optimal setting. This demonstrates that HEISIR adds value beyond simple score aggregation.

\begin{table}[htbp]
    \centering
    \resizebox{\columnwidth}{!}{
    \begin{tabular}{lcccc}
        \toprule
        combination & acc@1 & ndcg@5 & ndcg@10 & ndcg@20 \\
        \midrule
        \textit{ensemble effect} & 0.3445 & 0.2784 & 0.2651 & 0.2802 \\
        \small (- NV-Embed) & 0.3522 & 0.286 & 0.2715 & 0.2887 \\
        $\text{HEISIR}_{\scalebox{0.7}{Phi-3,GTE}}$ & 0.3582 & 0.2934 & 0.2757 & 0.2901 \\
        $\text{HEISIR}_{\scalebox{0.7}{Phi-3,LLM2Vec}}$ & 0.3865 & 0.3138 & 0.2936 & 0.3055 \\
        $\text{HEISIR}_{\scalebox{0.7}{Phi-3,OpenAI-large}}$ & \textbf{0.3950} & \textbf{0.3192} & \textbf{0.2992} & \textbf{0.3136} \\
        \bottomrule
    \end{tabular}
    }
    \caption{Comparison of Ensemble Effects}
    \label{tab:ensemble}
\end{table}

\subsection{Potential to Hybrid Search}
In section \ref{subsection:component}, we observed that progressively adding conversational components improved retrieval performance. Following this approach, we experimented with the addition of keywords to determine whether they enhance performance. Previous research on hybrid search methods report promising performance improvements by incorporating keyword to semantic search \cite{kuzi2020leveraging, lamer}. We explore addition of BM25 score into various HEISIR settings, with results shown in Table \ref{tab:bm25_comparison}. 

The performance results reveal interesting patterns in the integration of BM25 \cite{bm25} with various models. When combined with weaker models like Gemma and NV-Embed, BM25 provided slight improvements. However, its integration with stronger models (\( \text{HEISIR}_{\text{others}} \)) led to performance declines. This trend aligns with our previous observations in Section \ref{subsection:practical}. These findings suggest that BM25's simplistic relevance estimation may struggle to capture the complex semantic relationships present in conversational contexts.

\begin{table}[htbp]
\centering
\resizebox{\columnwidth}{!}{%
\begin{tabular}{lccccc}
\toprule
\textbf{Model} & \textbf{acc@1} & \textbf{acc@5} & \textbf{ndcg@10} & \textbf{ndcg@20} \\
\midrule
Only BM25 & 0.1465 & 0.3610 & 0.1200 & 0.1299 \\
\midrule
$\text{HEISIR}_{\scalebox{0.7}{Gemma, All-embeddings}}$ & 0.3048 & 0.5919 & 0.2315 & 0.2430 \\
+ BM25 score & \textbf{0.3105} & \textbf{0.5939} & \textbf{0.2346} & \textbf{0.2465} \\
\midrule
$\text{HEISIR}_{\scalebox{0.7}{All-LLMs,NV-Embed}}$  & 0.2806 & 0.5542 & 0.2107 & 0.2196 \\
+ BM25 score & \textbf{0.2912} & \textbf{0.5688} & \textbf{0.2189} & \textbf{0.2279} \\
\midrule
$\text{HEISIR}_{\scalebox{0.7}{others}}$ & \textbf{0.3822} & \textbf{0.6849} & \textbf{0.2886} & \textbf{0.3020} \\
+ BM25 score  & 0.3767 & 0.6791 & 0.2858 & 0.2997 \\
\bottomrule
\end{tabular}%
}
\caption{Metric Changes with and without BM25}
\label{tab:bm25_comparison}
\end{table}

\subsection{Optimization of HEISIR through weighted sum}

HEISIR currently calculates the final score by equally summing all component scores. However, components may contribute differently to sentence meaning and overall performance, suggesting potential benefits from a weighted sum approach. Our experiments using random search demonstrate clear potential for performance improvement through weight optimization, as shown in Table \ref{tab:weighted_sum}. These weights can be further refined for specific domains, potentially yielding even better results.

\begin{table}[t]
\centering
\resizebox{0.9\columnwidth}{!}{%
\small
\setlength{\tabcolsep}{6pt}
\begin{tabular}{@{}c*{4}{c}@{}}
\toprule
\multirow{3}{*}{\makecell{\textbf{Model}}} & \multicolumn{2}{c}{\textbf{Baseline}} & \multicolumn{2}{c}{\textbf{Weighted}} \\
\cmidrule(lr){2-3} \cmidrule(lr){4-5}
& acc@1 & ndcg@20 & acc@1 & ndcg@20 \\
\midrule
\makecell{OpenAI-large \\ + GPT-3.5-turbo} & 0.4085 & 0.3198 & \textbf{0.4179} & \textbf{0.3202} \\[6pt]
\cdashline{1-5}[0.5pt/1pt]
\\[-7pt]
\makecell{LLama-3 \\ + LLM2Vec} & 0.3902 & 0.3073 & \textbf{0.4050} & \textbf{0.3097} \\
\bottomrule
\end{tabular}%
}
\caption{Comparison of model performance with baseline and weighted sum approaches}
\label{tab:weighted_sum}
\end{table}

\subsection{Applications of Semantic Indexing}

Utilizing HEISIR for retrieval offers two key advantages beyond performance. First, it enhances interpretability: the highest-scoring SVOA indices provide structured, detailed insights into retrieval results. Second, HEISIR enables easy and effective result modification. Traditional retrieval systems often struggle to exclude unwanted results across semantically similar queries, but HEISIR overcomes this by allowing the removal of specific semantic indices responsible for undesired results,

Furthermore, the SV, SVO, and SVOA indices provide a guidance to user intent and conversation topics, aiding dialogue state tracking. Analysis of SI embeddings from OpenAI-large reveals clear clustering patterns that reproduce most intent labels in the USS dataset and encompass intents from other datasets (Table \ref{tab:intent_clusters}). This demonstrates the potential of HEISIR for automatic intent classification without fine-tuning. The potential for topic analysis using SVOA is further explored in Appendix \ref{apdx:potential_topic}.

\begin{table}[htbp]
\centering
\resizebox{\columnwidth}{!}{%
\renewcommand{\arraystretch}{1.3}%
\begin{tabular}{>{\Large\bfseries}l >{\Large}p{0.62\textwidth}}
\toprule
Intent Label & Representative Samples \\
\midrule
Rejection & Declines, denies, refuses, rejects \\
\addlinespace
Suggestion \& Planning & Suggests, wants, plans, advises, offers \\
\addlinespace
Expression \& Description \& Explanation & Expresses, describes, implies, explains \\
\addlinespace
Request & Requests, asks for, needs, seeks \\
\addlinespace
Preference & Likes, loves, enjoys, dislikes \\
\addlinespace
Acknowledgment \& Gratitude & Acknowledges, greets, thanks, appreciates \\
\addlinespace
Inquiry \& Curiosity & Inquires about, wants to know, wonders \\
\addlinespace
Mention \& Specification & States, specifies, confirms, clarifies \\
\addlinespace
Question & Asks, questions \\
\addlinespace
Mention & Mentions \\
\bottomrule
\end{tabular}%
}
\caption{Intent Clusters and Representative Samples}
\label{tab:intent_clusters}
\end{table}

\section{Related Work}
\textbf{Semantic Inverted Indexing for Retrieval}
Traditional retrieval methods streamline text searching with inverted indices \cite{zobel1998inverted, gormley2015elasticsearch, dey2024search}, while knowledge graph embeddings (KGEs) focus on revealing the innate structure of questions and answers \cite{wang2017knowledge, bordes2014open}. However, term-based approaches fail to capture context in multi-turn conversations, and KGEs struggle with representing implicit relationships \cite{ge2024knowledge}. There are very few researches that utilized SVO constituents as semantic indices, but were limited to simplistic SVO structure losing all semantic adjuncts \cite{burek2007svo, gao2009semantic}. Our research bridges these gaps with hierarchically expanded semantic indices, offering conversation-centric and detailed semantic representations than existing approaches.

\noindent \textbf{Semantic Search}
Semantic search models have shown capabilities in handling the semantic complexity in natural language queries. Encoder-based models, such as SBERT \cite{sbert},  DPR \cite{dpr} and SimCSE \cite{simcse}, have shown high performance upon their introduction but lacked scalability in general. Current research efforts, including SGPT \cite{sgpt}, RepLLaMA \cite{repllama}, leverage LLMs to address the limitation of encoder-based approaches. However, LLM-based models typically require substantial computational resources for training and deployment. In contrast, HEISIR can achieve high performance without the need for labeling and training.

\section{Conclusion}
In this paper, we introduced HEISIR, a novel framework for conversational data retrieval that reflects human sentence comprehension by capturing the syntactic hierarchy in natural language. HEISIR employs a 2-step extraction process to significantly improve the precision of semantic indices. Furthermore, by building semantic understanding from natural language syntax, HEISIR alleviates the need for extensive labeling and training.

Our experiments demonstrate that HEISIR consistently enhances retrieval performance across a variety of settings. Additionally, the inherent interpretability of the method offers significant advantages, making intent and topic analysis in dialogue datasets more accessible, thus providing a versatile solution for conversational AI systems. As chat-based services continue to grow rapidly, our findings have the potential to greatly enhance conversational data retrieval for both end-users and service providers.

\section*{Limitations}

\textbf{Multilingual}: HEISIR extracts quadruplets from messages based on the syntax of English, which is an isolating language. In isolating languages, word order plays a crucial role in semantic comprehension since each word typically contains only one morpheme. This suggests that HEISIR may encounter challenges when extracting SVOA quadruplets from agglutinative and fusional languages, where word order does not directly reflect the syntactic hierarchy.

\textbf{Expanding HEISIR to Document Data}:
Extending HEISIR from conversations to general documents presents several challenges. Conversational data have clear speakers as subjects in each utterance. However, documents often use passive voice or describe third-party actions, creating complex subject relationships that are difficult to parse. Furthermore, while conversations typically contain clear actions, documents tend to focus more on describing states and concepts, which makes SVOA analysis more challenging. This complexity necessitates advanced prompting and preprocessing techniques to accurately capture the subject and context. Additionally, it becomes important to account for avalent verbs in document retrieval, which are not considered in conversational data.

\textbf{Weight Optimization}: HEISIR currently employs simple summation to compute similarity scores. However, performance improvements were observed when assigning different weights to each conversational component. While uniform weight aggregation already outperforms current state-of-the-art retrieval models, further exploration is needed to optimize these weights for maximum efficiency.

\bibliography{main}

\newpage
\appendix
\section{Prompts for HEISIR}
\label{apdx:prompts}
We detail the prompts used for our method. All prompts use \{\{\$variable\}\} as placeholders for external variables. While the prompts shown here are designed for GPT-3.5-turbo, we adapted them for other LLM models by adding predefined tokens according to each model's specific template. Each prompt consists of four rows in the prompt table: common LLM parameters, system prompt, 5-shot examples generated using GPT-4o (omitted in this appendix for brevity), and user message with input placeholders.

\subsection{Prompt for Hierarchical Triplets Formulation}

\begin{table}[H]
\setlength{\tabcolsep}{2pt}
\begin{tabular}{|p{0.97\columnwidth}|}
\hline
\scriptsize\textbf{Temperature}: 0.0 \newline
\scriptsize\textbf{max\_tokens}: 1024
\\ \hline
\vspace{-1.0em}
\scriptsize
\textbf{System:}
\vspace{1pt}

\tiny
You must extract all $[$\$information triplet\$$]$ in the message, given the context and role, subject to the following conditions:
\begin{itemize}[leftmargin=*,itemsep=0pt,parsep=0pt,topsep=0pt,partopsep=0pt]
\tiny
    \item $[$\$information triplet\$$]$ should include anything you need to understand the message and the role, such as the role's emotions, the topic of conversation, the intent of the message and etc.
    \item You need to extract an $[$\$information triplet\$$]$ for a new message according to the following instructions.
\end{itemize}
\tiny
Structural Constraints of $[$\$information triplet\$$]$:
\begin{itemize}[leftmargin=*,itemsep=0pt,parsep=0pt,topsep=0pt,partopsep=0pt]
\tiny
    \item The structure of $[$\$information triplet\$$]$ : $[$\$subject\$$]$ $[$\$common verb\$$]$ $[$\$target content\$$]$
    \item $[$\$subject\$$]$ : Must be the \{\{\$role\}\}.
    \item $[$\$common verb\$$]$:
    \begin{itemize}[itemsep=0pt,parsep=0pt,topsep=0pt,partopsep=0pt]
    \tiny
        \item Must Use a singular present tense verb (e.g., "says", "asks", "wants to", "inquires about", "looks into") to describe the \{\{\$role\}\}'s action or intention.
        \item If you need to use the negative form, write it as not in front of the common verb. However, use negative forms only when they are essential to understanding the sentence. 
    \end{itemize}
    \item $[$\$target content\$$]$:
    \begin{itemize}[itemsep=0pt,parsep=0pt,topsep=0pt,partopsep=0pt]
    \tiny
        \item Must be a noun phrase of 3 characters or less
        \item If $[$\$target content\$$]$ contains content too specific to be generalized, such as a person's name, webpage address, or code, generalize it by using general expressions such as person, friend, url, website, code and etc.
        \item Should contain only one content; if multiple noun phrases or contents are needed, separate them into individual $[$\$information triplet\$$]$ items.
    \end{itemize}
\end{itemize}
\tiny
How to Construct an Information Triplet:
\begin{enumerate}[leftmargin=*,itemsep=0pt,parsep=0pt,topsep=0pt,partopsep=0pt]
\tiny
    \item You receive the $[$\$conversation context\$$]$ along with the $[$\$message\$$]$ you need to analyze as input.
    \item Review the [\$conversation context\$] to understand the nuance and content of the $[$\$message\$$]$. However, the [\$conversation context\$] should only be used to understand the message and should not be used to extract the $[$\$information triplet\$$]$.
    \item Extract all the $[$\$information triplet\$$]$ that can be obtained from the message in the form of a JSON.
    \item The JSON format is as follows:
\end{enumerate}

\vspace{0.5em}

\begin{varwidth}{\linewidth}
\begin{verbatim}
{
  "information_triplet":
  [
    {"{{$role}} [$common verb$]": [$target content$]},
    {"{{$role}} [$common verb$]": [$target content$]},
    {"{{$role}} [$common verb$]": [$target content$]},
    ...  
  ]
}
\end{verbatim}
\end{varwidth}
\\ \hline
\scriptsize
\textbf{Few-shot examples}
\\ \hline
\scriptsize
\textbf{User:}
\vspace{1pt}

\tiny
$[$\$conversation context\$$]$
\vspace{1pt}

\tiny \{\{\$context\}\}
\\

\tiny
$[$\$message\$$]$
\vspace{1pt}

\tiny
\{\{\$role\}\}: \{\{\$message\}\}
\\
\tiny
Extract as much $[$\$information triplet\$$]$ as possible from the message while maintaining all of the above conditions.
We recommend extracting between 5 and 20 pieces of $[$\$information triplet\$$]$, depending on the length of the sentence.
The key of the information\_triplet you create should start with \{\{\$role\}\}.
\\
\tiny
Answer:
\\
\hline
\end{tabular}
\caption{Prompt and parameters for Hierarachical Triplets Expansion}
\label{tab:prompt}
\end{table}

\subsection{Adjunct Augmentation}

\begin{table}[H]
\setlength{\tabcolsep}{2pt}
\centering
\begin{tabular}{|p{0.97\columnwidth}|}
\hline
\scriptsize\textbf{Temperature}: 0.0 \newline
\scriptsize\textbf{max\_tokens}: 1024
\\ \hline

\scriptsize\textbf{System:}
\vspace{1pt}

\tiny
You will be provided with \$conversation context\$, \$message\$ and \$information list\$. 
\$message\$ is a real conversation message from \{\{\$role\}\} that you need to analyze. 
\$information list\$ is information extracted from \$message\$, consisting of \$subject\$, \$verb\$, and \$target content\$.

You need to elaborate on the \$detail\$ of \$target content\$ according to the following instructions.
\begin{itemize}[leftmargin=*,itemsep=0pt,parsep=0pt,topsep=0pt,partopsep=0pt]
\tiny
    \item Prepositions must be actively used to describe the \$detail\$ of the target content.
    \item If the value placed in \$detail\$ is vague, such as a pronoun, please use a specific noun to make the meaning clear.
    \item Here are three recommended strategies for elaborating on the \$detail\$:
    \begin{enumerate}[leftmargin=*,itemsep=0pt,parsep=0pt,topsep=0pt,partopsep=0pt]
    \tiny
        \item \textbf{Detailed Content and theme of Discussion}
        
        This category involves prepositions that help specify the subject of communication, the object of emotions or actions, and the theme of a conversation.
        For example
        \begin{itemize}[leftmargin=*,itemsep=0pt,parsep=0pt,topsep=0pt,partopsep=0pt]
        \tiny
            \item user expresses anger: to assistant.
            \item user asks questions: about climate change.
            \item assistant maintains a professional attitude: towards the user's queries.
            \item assistant offers advice: regarding data protection.
            \item user expresses confusion: over the assistant's instructions.
            \item user seeks clarification: with respect to the subscription plans.
            \item assistant invests effort: in improving the user interface.
        \end{itemize}
        \item \textbf{Reasons or Causations}
        
        This set of phrases explains the cause or reason behind an action or situation. It answers the "why" of a scenario.
        For example
        \begin{itemize}[leftmargin=*,itemsep=0pt,parsep=0pt,topsep=0pt,partopsep=0pt]
        \tiny
            \item assistant pauses the service: because of maintenance needs.
            \item user misses the deadline: due to a technical glitch.
            \item user returns the product: owing to a manufacturing defect.
            \item assistant improves response time: thanks to the user's constructive feedback.
        \end{itemize}
        \item \textbf{Conditions or accompanying Circumstances}
        
        These prepositions are used to describe the conditions under which something happens or the context that accompanies an action.
        For example
        \begin{itemize}[leftmargin=*,itemsep=0pt,parsep=0pt,topsep=0pt,partopsep=0pt]
        \tiny
            \item user solves problems: with patience.
            \item assistant writes articles: for users.
            \item user reads the manual: over the weekend.
            \item user leaves feedback: on the website.
            \item user places the report: under the book.
        \end{itemize}
        \item \textbf{In cases where no specific details can be included}
        
        If a sentence structure consisting of \{\{\$role\}\} \$verb\$ \$target content\$ \$detail\$ is impossible or adding detailed information is not meaningful, set "no information" as the value for \$detail\$.
        For example
        \begin{itemize}[leftmargin=*,itemsep=0pt,parsep=0pt,topsep=0pt,partopsep=0pt]
        \tiny
            \item user mentions Christmas: no information.
        \end{itemize}
    \end{enumerate}
    \item The answer should be written in JSON format as follows:
\end{itemize}

\vspace{0.5em}
\begin{varwidth}{\linewidth}
\begin{verbatim}
{
  "detailed_information":
  [
    {"{{$role}} [$verb$] [$target content$]": "[$detail$]"},
    {"{{$role}} [$verb$] [$target content$]": "[$detail$]"},
    ...  
  ]
}
\end{verbatim}
\end{varwidth}
\\ \hline
\scriptsize
\textbf{Few-shot examples}
\\ \hline
\scriptsize
\textbf{User:}
\vspace{1pt}

\tiny
$[$\$conversation context\$$]$
\vspace{1pt}

\tiny \{\{\$context\}\}
\\

\tiny
$[$\$message\$$]$
\vspace{1pt}

\tiny
\{\{\$role\}\}: \{\{\$message\}\}
\\

\tiny
$[$\$information list\$$]$
\vspace{1pt}

\tiny
\{\{\$info\_list\}\}
\\

\tiny
Choose the best of the three strategies above and write a 2-3 word answer that clarifies the content of the sentence.
Do not include the content of the sentence in your answer, but start with the preposition.
The entire contents of [\$Information list\$] should be used as the key for detailed\_information without any changes at all.
\\
\tiny
Answer:
\\
\hline
\end{tabular}
\caption{Prompt and parameters for Detailed Description Augmentation}
\label{tab:prompt2}
\end{table}


\section{All results}
\label{apdx:results}
Table [\ref{tab:gemma_all_result}-\ref{tab:haiku_all_result}] displays performance for key component combinations across LLM and Embedding models. 'p' and 'r' represent precision and recall, respectively. The highest value per metric is in bold, with the second-highest underlined. 

\input{tables/gemma}
\input{tables/phi}
\input{tables/llama}
\input{tables/qwen}
\input{tables/gpt}
\input{tables/haiku}
\input{tables/base}

Table \ref{tab:base} shows the baseline performance of all embedding models when using only conversation and messages.

\section{Intent \& Topic Analysis}
\label{apdx:potential_topic}
In the main text, we observed intent clustering using SV. SVOA or SVO can reveal specific conversation topics. Table \ref{apdx:potential_topic} shows the results of clustering into 15 groups. Examining representative topics for each cluster demonstrates that we can recover most major conversation themes from our dataset, along with related topics.

This analysis uses a basic K-means algorithm, showcasing the potential of Semantic indices for intent and topic analysis. While the current approach is fundamental, it effectively captures and categorizes conversational content. We anticipate that more advanced methods could yield even more robust and nuanced models, opening up various possibilities for future research in natural language processing and conversational AI.

\section{Reproducing}
\label{apdx:reproducing}
For all experiments, we utilized a single NVIDIA H100 80GB HBM3 GPU. Our study employed various models from Hugging Face and API. The embedding models consisted of Alibaba-NLP/gte-large-en-v1.5 (GTE), nomic-ai/nomic-embed-text-v1 (Nomic), nvidia/NV-Embed-v1 (NV-Embed), and McGill-NLP/LLM2Vec-Meta-Llama-3-8B-Instruct-mntp-supervised (LLM2Vec). Language models included google/gemma-2b-it (Gemma), microsoft/Phi-3-mini-128k-instruct (Phi-3), meta-llama/Meta-Llama-3-8B-Instruct (LLama-3), Qwen/Qwen2-7B-Instruct (Qwen-2), GPT-3.5-turbo-0125, and claude-3-haiku-20240307. In the fine-tuning process, train/validation datasets were created like test data, using only train data and excluding DICES due to data unavailability. Fine-tuning ran for 10 epochs. For LLM2Vec, we used LoRA (r=16, alpha=32) with a 1e-4 learning rate; other models used 1e-5. All code and data are accessible in the supplementary materials.

\begin{table}[htbp]
\centering
\resizebox{\columnwidth}{!}{%
\renewcommand{\arraystretch}{1.3}%
\begin{tabular}{>{\Large\bfseries}l >{\Large}p{0.62\textwidth}}
\toprule
Cluster & Representative Samples \\
\midrule
Restaurants \& Food & Asks restaurant, mentions food, asks food, asks music at restaurant \\
\addlinespace
Movies & Likes movie, asks movie, mentions movie, acknowledges movie \\
\addlinespace
Well-being \& Health & Asks question about well-being, inquires well-being, asks opinion \\
\addlinespace
Reservations & Specifies number of people, confirms booking, asks booking \\
\addlinespace
Family \& Pets & Mentions family, mentions work, mentions kids, mentions dog \\
\addlinespace
Sports \& Hobbies & Asks sports, asks interest in sports, mentions football, plays games \\
\addlinespace
Pricing \& Admission & Asks fee for entrance, asks method of payment, mentions price \\
\addlinespace
Conversation Closure & Declines help, declines offer, says goodbye, ends conversation \\
\addlinespace
Miscellaneous & Has kids, greets morning, feels tired, watches TV \\
\addlinespace
Location Inquiries & Asks location, asks hotel, mentions hotel, asks duration of stay \\
\addlinespace
Interest in Others & Asks you, greets person with question, asks activity, asks today \\
\addlinespace
Greetings & Greets person, expresses gratitude, acknowledges response \\
\addlinespace
Music \& Pets & Likes music, mentions hobbies, has dog, enjoys music \\
\addlinespace
Recommendations & Thinks better, asks for recommendations, suggests change of topic \\
\addlinespace
Asking Opinions & Expresses opinion, states opinion, expresses frustration \\
\bottomrule
\end{tabular}%
}
\caption{Intent Clusters and Representative Samples from Data}
\label{tab:intent_clusters_data}
\end{table}

\section{Impact of Context}

\begin{table}[t]
\centering
\resizebox{\columnwidth}{!}{%
\scriptsize
\setlength{\tabcolsep}{2pt}
\renewcommand{\arraystretch}{0.9}
\begin{tabular}{cccccc}
\toprule
\multirow{2}{*}{\textbf{Model}} & \multirow{2}{*}{\textbf{Context}} & \multicolumn{4}{c}{\textbf{Metric}} \\
\cmidrule(l){3-6}
& & \textbf{acc@1} & \textbf{ndcg@5} & \textbf{ndcg@10} & \textbf{ndcg@20} \\
\midrule
\multirow{2}{*}{\makecell[l]{$\text{HEISIR}_{\scalebox{0.6}{\begin{tabular}[c]{@{}c@{}}GPT-3.5-turbo,\\All-embeddings\end{tabular}}}$}} & x & 0.3569 & 0.2869 & 0.2691 & 0.2813 \\
& o & \textbf{0.3683} & \textbf{0.2959} & \textbf{0.2772} & \textbf{0.2902} \\
\bottomrule
\end{tabular}%
}
\caption{Metric Changes with and without context}
\label{tab:context}
\end{table}


We select GPT-3.5-turbo as the LLM as it is the most accessible and widely model used in industry. Including previous context for semantic indices construction shows minor improvement in retrieval performance, as shown in Table \ref{tab:context}.

However, in terms of interpretability, a thorough analysis of SVOA indices reveals that SVOA quadruplets extracted without previous context fail to capture relational dependencies between words. Since HEISIR avoids the use of pronouns to ensure semantic completeness, this lack of context makes it difficult to extract accurate information. Consequently, extracting SVOA quadruplets from context-less messages not only reduces the precision of the semantic indices but also results in fewer indices being generated. For future applications beyond retrieval tasks, incorporating previous context will be necessary. 

\section{Dataset License and Disclaimer}
The datasets used in this study are subject to the following licenses: BAD \cite{databad} and DICES \cite{datadices} (CC BY 4.0), DailyDialog \cite{datadaily} (CC BY-NC-SA 4.0), PILD \cite{datapild} (MIT license), USS \cite{datauss} (individual licenses), ReDial (CC-BY-4.0), MWOZ (MIT License), and SGD (CC-BY-SA-4.0). We strictly adhered to these licenses and confirm that no commercial use was made of any dataset in this research. While BAD and DICES datasets contain some inherently harmful content, we used these datasets solely as search targets and emphasize that our paper does not include any harmful content in its presented material.

\end{document}

%% file: macros.tex
\newcommand{\todoc}[2]{{\textcolor{#1}{\textbf{#2}}}}
\newcommand{\todoblue}[1]{\todoc{blue}{\textbf{#1}}}
\newcommand{\todored}[1]{\todoc{red}{#1}}
\newcommand{\todoorange}[1]{\todoc{orange}{#1}}
\newcommand{\todopurple}[1]{\todoc{purple}{#1}}
\newcommand{\todogreen}[1]{\todoc{green}{#1}}

\newcommand{\yohan}[1]{\todopurple{\textbf{yohan:} #1}}

\newcommand\Tstrut{\rule{0pt}{2.2ex}}       
\newcommand\Bstrut{\rule[-0.6ex]{0pt}{0pt}} 
\newcommand{\TBstrut}{\Tstrut\Bstrut} 

%% file: tables/gemma.tex
\begin{table*}[htbp]
\centering
\scriptsize
\setlength{\tabcolsep}{4pt}
\setlength{\tabcolsep}{3pt} 
\renewcommand{\arraystretch}{0.95} 
\begin{tabular}{lcccccccccccccc}
\toprule
embedding & combination & acc@1 & acc@5 & p@5 & p@10 & r@5 & r@10 & ndcg@10 & ndcg@20 & mrr@10 & mrr@20 & map@10 & map@20 \\
\midrule
\multirow{6}{*}{\centering GTE}
& sv & 0.0229 & 0.0717 & 0.0159 & 0.0133 & 0.0058 & 0.0096 & 0.0158 & 0.0168 & 0.0441 & 0.0481 & 0.0059 & 0.0050 \\
& sv \_svo & 0.1311 & 0.3353 & 0.1020 & 0.0873 & 0.0440 & 0.0724 & 0.1045 & 0.1136 & 0.2189 & 0.2277 & 0.0518 & 0.0494 \\
& sv \_svo\_svoa & 0.1899 & 0.4467 & 0.1492 & 0.1271 & 0.0709 & 0.1127 & 0.1552 & 0.1688 & 0.2983 & 0.3062 & 0.0831 & 0.0820 \\
& svoa\_conv\_msg & \textbf{0.3253} & \textbf{0.6252} & \textbf{0.2424} & \textbf{0.1990} & \textbf{0.1179} & \textbf{0.1805} & \textbf{0.2504} & \textbf{0.2643} & \textbf{0.4514} & \textbf{0.4577} & \textbf{0.1486} & \textbf{0.1450} \\
& svo\_svoa\_conv\_msg & \underline{0.3131} & \underline{0.6181} & \underline{0.2400} & \underline{0.1969} & \underline{0.1150} & \underline{0.1776} & \underline{0.2465} & \underline{0.2594} & \underline{0.4425} & \underline{0.4489} & \underline{0.1458} & \underline{0.1414} \\
& sv \_svo\_svoa\_conv\_msg & 0.3096 & 0.6124 & 0.2350 & 0.1929 & 0.1120 & 0.1729 & 0.2415 & 0.2541 & 0.4374 & 0.4437 & 0.1423 & 0.1378 \\
\midrule
\multirow{6}{*}{\centering Nomic}
& sv & 0.0157 & 0.0520 & 0.0115 & 0.0106 & 0.0041 & 0.0077 & 0.0120 & 0.0132 & 0.0325 & 0.0363 & 0.0044 & 0.0037 \\
& sv \_svo & 0.1245 & 0.3056 & 0.0972 & 0.0827 & 0.0416 & 0.0676 & 0.0984 & 0.1065 & 0.2030 & 0.2114 & 0.0497 & 0.0473 \\
& sv \_svo\_svoa & 0.1799 & 0.4264 & 0.1459 & 0.1232 & 0.0674 & 0.1083 & 0.1497 & 0.1594 & 0.2874 & 0.2952 & 0.0806 & 0.0777 \\
& svoa\_conv\_msg & \textbf{0.2959} & \textbf{0.5910} & \underline{0.2234} & \textbf{0.1806} & \underline{0.1088} & \textbf{0.1650} & \textbf{0.2295} & \textbf{0.2421} & \textbf{0.4197} & \textbf{0.4265} & \textbf{0.1354} & \textbf{0.1319} \\
& svo\_svoa\_conv\_msg & \underline{0.2939} & \underline{0.5855} & \textbf{0.2246} & \underline{0.1803} & \textbf{0.1089} & \underline{0.1628} & \underline{0.2282} & \underline{0.2400} & \underline{0.4179} & \underline{0.4245} & \underline{0.1344} & \underline{0.1299} \\
& sv \_svo\_svoa\_conv\_msg & 0.2919 & 0.5793 & 0.2203 & 0.1781 & 0.1064 & 0.1606 & 0.2244 & 0.2344 & 0.4117 & 0.4182 & 0.1317 & 0.1263 \\
\midrule
\multirow{6}{*}{\centering NV-Embed}
& sv & 0.0183 & 0.0620 & 0.0137 & 0.0119 & 0.0049 & 0.0083 & 0.0137 & 0.0157 & 0.0376 & 0.0425 & 0.0050 & 0.0043 \\
& sv \_svo & 0.1282 & 0.3196 & 0.0979 & 0.0824 & 0.0418 & 0.0672 & 0.0992 & 0.1052 & 0.2114 & 0.2192 & 0.0495 & 0.0464 \\
& sv \_svo\_svoa & 0.1879 & 0.4247 & 0.1403 & 0.1170 & 0.0652 & 0.1016 & 0.1442 & 0.1544 & 0.2878 & 0.2956 & 0.0769 & 0.0744 \\
& svoa\_conv\_msg & 0.2339 & 0.4664 & 0.1616 & 0.1297 & 0.0754 & 0.1143 & 0.1656 & 0.1723 & 0.3326 & 0.3406 & 0.0922 & 0.0868 \\
& svo\_svoa\_conv\_msg & \underline{0.2482} & \textbf{0.4976} & \textbf{0.1754} & \textbf{0.1419} & \textbf{0.0819} & \textbf{0.1237} & \textbf{0.1793} & \underline{0.1870} & \underline{0.3536} & \underline{0.3607} & \textbf{0.1008} & \underline{0.0955} \\
& sv \_svo\_svoa\_conv\_msg & \textbf{0.2505} & \underline{0.4970} & \underline{0.1745} & \underline{0.1416} & \underline{0.0808} & \underline{0.1236} & \underline{0.1791} & \textbf{0.1871} & \textbf{0.3556} & \textbf{0.3629} & \underline{0.1004} & \textbf{0.0953} \\
\midrule
\multirow{6}{*}{\centering LLM2Vec}
& sv & 0.0240 & 0.0686 & 0.0155 & 0.0143 & 0.0058 & 0.0104 & 0.0166 & 0.0184 & 0.0447 & 0.0496 & 0.0062 & 0.0054 \\
& sv \_svo & 0.1394 & 0.3585 & 0.1136 & 0.0969 & 0.0497 & 0.0803 & 0.1151 & 0.1247 & 0.2326 & 0.2411 & 0.0585 & 0.0562 \\
& sv \_svo\_svoa & 0.1974 & 0.4622 & 0.1578 & 0.1317 & 0.0746 & 0.1171 & 0.1611 & 0.1739 & 0.3078 & 0.3153 & 0.0877 & 0.0863 \\
& svoa\_conv\_msg & 0.3082 & 0.6130 & 0.2338 & 0.1909 & 0.1134 & \underline{0.1740} & 0.2410 & 0.2538 & 0.4361 & 0.4422 & 0.1419 & 0.1381 \\
& svo\_svoa\_conv\_msg & \textbf{0.3165} & \underline{0.6195} & \underline{0.2378} & \textbf{0.1930} & \underline{0.1150} & \textbf{0.1743} & \textbf{0.2439} & \textbf{0.2563} & \textbf{0.4421} & \textbf{0.4485} & \textbf{0.1443} & \textbf{0.1397} \\
& sv \_svo\_svoa\_conv\_msg & \underline{0.3136} & \textbf{0.6198} & \textbf{0.2387} & \underline{0.1923} & \textbf{0.1151} & 0.1737 & \underline{0.2428} & \underline{0.2543} & \underline{0.4410} & \underline{0.4474} & \underline{0.1436} & \underline{0.1387} \\
\midrule
\multirow{6}{*}{\centering OpenAI-small}
& sv & 0.0171 & 0.0628 & 0.0138 & 0.0123 & 0.0050 & 0.0091 & 0.0139 & 0.0158 & 0.0376 & 0.0423 & 0.0049 & 0.0043 \\
& sv \_svo & 0.1280 & 0.3376 & 0.1068 & 0.0909 & 0.0474 & 0.0757 & 0.1079 & 0.1178 & 0.2174 & 0.2263 & 0.0549 & 0.0529 \\
& sv \_svo\_svoa & 0.1962 & 0.4562 & 0.1546 & 0.1308 & 0.0728 & 0.1171 & 0.1599 & 0.1711 & 0.3059 & 0.3134 & 0.0867 & 0.0845 \\
& svoa\_conv\_msg & \textbf{0.3253} & \textbf{0.6190} & \textbf{0.2398} & \textbf{0.1976} & \textbf{0.1173} & \textbf{0.1813} & \textbf{0.2503} & \textbf{0.2656} & \textbf{0.4508} & \textbf{0.4573} & \textbf{0.1495} & \textbf{0.1469} \\
& svo\_svoa\_conv\_msg & \underline{0.3242} & \underline{0.6178} & \underline{0.2392} & \underline{0.1957} & \underline{0.1166} & \underline{0.1779} & \underline{0.2477} & \underline{0.2614} & \underline{0.4485} & \underline{0.4549} & \underline{0.1476} & \underline{0.1442} \\
& sv \_svo\_svoa\_conv\_msg & 0.3208 & 0.6084 & 0.2355 & 0.1925 & 0.1150 & 0.1752 & 0.2440 & 0.2574 & 0.4448 & 0.4511 & 0.1451 & 0.1416 \\
\midrule
\multirow{6}{*}{\centering OpenAI-large}
& sv & 0.0163 & 0.0603 & 0.0135 & 0.0119 & 0.0054 & 0.0089 & 0.0136 & 0.0152 & 0.0360 & 0.0405 & 0.0049 & 0.0042 \\
& sv \_svo & 0.1325 & 0.3399 & 0.1062 & 0.0918 & 0.0472 & 0.0764 & 0.1091 & 0.1183 & 0.2218 & 0.2304 & 0.0554 & 0.0532 \\
& sv \_svo\_svoa & 0.1962 & 0.4582 & 0.1545 & 0.1313 & 0.0734 & 0.1176 & 0.1604 & 0.1727 & 0.3078 & 0.3157 & 0.0864 & 0.0848 \\
& svoa\_conv\_msg & \textbf{0.3508} & \textbf{0.6590} & \textbf{0.2587} & \textbf{0.2130} & \textbf{0.1271} & \textbf{0.1941} & \textbf{0.2694} & \textbf{0.2851} & \textbf{0.4798} & \textbf{0.4856} & \textbf{0.1627} & \textbf{0.1600} \\
& svo\_svoa\_conv\_msg & \underline{0.3473} & \underline{0.6387} & \underline{0.2527} & \underline{0.2071} & \underline{0.1226} & \underline{0.1875} & \underline{0.2620} & \underline{0.2756} & \underline{0.4714} & \underline{0.4773} & \underline{0.1575} & \underline{0.1536} \\
& sv \_svo\_svoa\_conv\_msg & 0.3422 & 0.6347 & 0.2483 & 0.2031 & 0.1202 & 0.1836 & 0.2574 & 0.2705 & 0.4656 & 0.4714 & 0.1541 & 0.1502 \\
\bottomrule
\end{tabular}
\caption{Results for Different Embeddings and \textbf{Gemma}}
\label{tab:gemma_all_result}
\end{table*}

%% file: tables/phi.tex
\begin{table*}[htbp]
\centering
\scriptsize
\setlength{\tabcolsep}{4pt}
\setlength{\tabcolsep}{3pt} 
\renewcommand{\arraystretch}{0.95} 
\begin{tabular}{lcccccccccccccc}
\toprule
embedding & combination & acc@1 & acc@5 & p@5 & p@10 & r@5 & r@10 & ndcg@10 & ndcg@20 & mrr@10 & mrr@20 & map@10 & map@20 \\
\midrule
\multirow{6}{*}{\centering GTE}
& sv & 0.0374 & 0.1163 & 0.0280 & 0.0241 & 0.0113 & 0.0189 & 0.0286 & 0.0318 & 0.0728 & 0.0785 & 0.0116 & 0.0106 \\
& sv\_svo & 0.2396 & 0.5330 & 0.1864 & 0.1564 & 0.0879 & 0.1390 & 0.1920 & 0.2053 & 0.3642 & 0.3709 & 0.1065 & 0.1047 \\
& sv\_svo\_svoa & 0.2816 & 0.5698 & 0.2107 & 0.1748 & 0.1015 & 0.1577 & 0.2187 & 0.2336 & 0.4033 & 0.4099 & 0.1267 & 0.1248 \\
& svoa\_conv\_msg & 0.3476 & 0.6587 & 0.2638 & 0.2158 & 0.1270 & 0.1949 & 0.2716 & 0.2856 & 0.4805 & 0.4861 & 0.1643 & 0.1603 \\
& svo\_svoa\_conv\_msg & \textbf{0.3610} & \underline{0.6621} & \underline{0.2660} & \textbf{0.2191} & \underline{0.1280} & \textbf{0.1972} & \textbf{0.2760} & \textbf{0.2905} & \underline{0.4877} & \underline{0.4933} & \textbf{0.1680} & \textbf{0.1644} \\
& sv\_svo\_svoa\_conv\_msg & \underline{0.3582} & \textbf{0.6695} & \textbf{0.2677} & \underline{0.2184} & \textbf{0.1286} & \underline{0.1966} & \underline{0.2757} & \underline{0.2901} & \textbf{0.4886} & \textbf{0.4947} & \underline{0.1677} & \underline{0.1641} \\
\midrule
\multirow{6}{*}{\centering Nomic}
& sv & 0.0386 & 0.1060 & 0.0262 & 0.0214 & 0.0105 & 0.0170 & 0.0265 & 0.0281 & 0.0693 & 0.0738 & 0.0111 & 0.0099 \\
& sv\_svo & 0.2508 & 0.5079 & 0.1847 & 0.1544 & 0.0849 & 0.1343 & 0.1900 & 0.1995 & 0.3603 & 0.3673 & 0.1076 & 0.1034 \\
& sv\_svo\_svoa & 0.2913 & 0.5641 & 0.2163 & 0.1773 & 0.1020 & 0.1563 & 0.2216 & 0.2335 & 0.4077 & 0.4148 & 0.1306 & 0.1265 \\
& svoa\_conv\_msg & 0.3416 & 0.6264 & 0.2464 & 0.2022 & 0.1199 & 0.1841 & 0.2566 & 0.2701 & 0.4637 & 0.4700 & 0.1541 & 0.1505 \\
& svo\_svoa\_conv\_msg & \underline{0.3573} & \underline{0.6484} & \underline{0.2600} & \underline{0.2121} & \underline{0.1247} & \textbf{0.1919} & \underline{0.2688} & \underline{0.2817} & \underline{0.4815} & \underline{0.4880} & \underline{0.1635} & \underline{0.1590} \\
& sv\_svo\_svoa\_conv\_msg & \textbf{0.3602} & \textbf{0.6590} & \textbf{0.2635} & \textbf{0.2138} & \textbf{0.1257} & \underline{0.1916} & \textbf{0.2707} & \textbf{0.2839} & \textbf{0.4858} & \textbf{0.4920} & \textbf{0.1647} & \textbf{0.1603} \\
\midrule
\multirow{6}{*}{\centering NV-Embed}
& sv & 0.0371 & 0.1200 & 0.0297 & 0.0245 & 0.0115 & 0.0187 & 0.0287 & 0.0316 & 0.0736 & 0.0794 & 0.0116 & 0.0104 \\
& sv\_svo & 0.2616 & 0.5330 & 0.1907 & 0.1572 & 0.0878 & 0.1359 & 0.1949 & 0.2043 & 0.3750 & 0.3819 & 0.1097 & 0.1052 \\
& sv\_svo\_svoa & \textbf{0.2933} & \underline{0.5601} & \underline{0.2089} & \underline{0.1720} & \underline{0.0996} & \textbf{0.1534} & \underline{0.2168} & \textbf{0.2277} & \textbf{0.4064} & \textbf{0.4134} & \underline{0.1259} & \textbf{0.1219} \\
& svoa\_conv\_msg & 0.2465 & 0.4990 & 0.1776 & 0.1447 & 0.0827 & 0.1269 & 0.1818 & 0.1885 & 0.3520 & 0.3589 & 0.1027 & 0.0973 \\
& svo\_svoa\_conv\_msg & 0.2773 & 0.5541 & 0.2069 & 0.1702 & 0.0967 & 0.1497 & 0.2124 & 0.2210 & 0.3950 & 0.4011 & 0.1233 & 0.1182 \\
& sv\_svo\_svoa\_conv\_msg & \underline{0.2893} & \textbf{0.5630} & \textbf{0.2129} & \textbf{0.1739} & \textbf{0.0997} & \underline{0.1528} & \textbf{0.2180} & \underline{0.2266} & \underline{0.4053} & \underline{0.4117} & \textbf{0.1275} & \underline{0.1217} \\
\midrule
\multirow{6}{*}{\centering LLM2Vec}
& sv & 0.0537 & 0.1557 & 0.0385 & 0.0330 & 0.0149 & 0.0251 & 0.0390 & 0.0421 & 0.0986 & 0.1051 & 0.0162 & 0.0143 \\
& sv\_svo & 0.2862 & 0.5781 & 0.2155 & 0.1781 & 0.1009 & 0.1567 & 0.2203 & 0.2332 & 0.4101 & 0.4172 & 0.1272 & 0.1237 \\
& sv\_svo\_svoa & 0.3193 & 0.6073 & 0.2378 & 0.1930 & 0.1136 & 0.1725 & 0.2430 & 0.2568 & 0.4405 & 0.4470 & 0.1451 & 0.1421 \\
& svoa\_conv\_msg & 0.3579 & 0.6552 & 0.2610 & 0.2124 & 0.1263 & 0.1938 & 0.2699 & 0.2833 & 0.4838 & 0.4897 & 0.1627 & 0.1584 \\
& svo\_svoa\_conv\_msg & \underline{0.3790} & \underline{0.6904} & \underline{0.2817} & \underline{0.2273} & \underline{0.1351} & \underline{0.2044} & \underline{0.2880} & \underline{0.2998} & \underline{0.5081} & \underline{0.5136} & \underline{0.1769} & \underline{0.1712} \\
& sv\_svo\_svoa\_conv\_msg & \textbf{0.3865} & \textbf{0.6967} & \textbf{0.2860} & \textbf{0.2321} & \textbf{0.1368} & \textbf{0.2078} & \textbf{0.2936} & \textbf{0.3055} & \textbf{0.5166} & \textbf{0.5219} & \textbf{0.1807} & \textbf{0.1749} \\
\midrule
\multirow{6}{*}{\centering OpenAI-small}
& sv & 0.0497 & 0.1400 & 0.0356 & 0.0294 & 0.0141 & 0.0225 & 0.0353 & 0.0378 & 0.0896 & 0.0955 & 0.0147 & 0.0131 \\
& sv\_svo & 0.2625 & 0.5410 & 0.1964 & 0.1636 & 0.0918 & 0.1453 & 0.2035 & 0.2163 & 0.3828 & 0.3904 & 0.1161 & 0.1130 \\
& sv\_svo\_svoa & 0.3022 & 0.5844 & 0.2233 & 0.1841 & 0.1066 & 0.1650 & 0.2314 & 0.2451 & 0.4239 & 0.4310 & 0.1362 & 0.1333 \\
& svoa\_conv\_msg & 0.3653 & 0.6550 & 0.2627 & 0.2167 & 0.1285 & 0.1986 & 0.2752 & 0.2915 & 0.4893 & 0.4956 & 0.1677 & 0.1654 \\
& svo\_svoa\_conv\_msg & \underline{0.3807} & \underline{0.6724} & \underline{0.2736} & \underline{0.2246} & \underline{0.1333} & \underline{0.2032} & \underline{0.2849} & \underline{0.2989} & \underline{0.5054} & \underline{0.5112} & \underline{0.1749} & \underline{0.1713} \\
& sv\_svo\_svoa\_conv\_msg & \textbf{0.3893} & \textbf{0.6821} & \textbf{0.2785} & \textbf{0.2283} & \textbf{0.1347} & \textbf{0.2060} & \textbf{0.2898} & \textbf{0.3034} & \textbf{0.5147} & \textbf{0.5201} & \textbf{0.1783} & \textbf{0.1740} \\
\midrule
\multirow{6}{*}{\centering OpenAI-large}
& sv & 0.0457 & 0.1328 & 0.0324 & 0.0284 & 0.0127 & 0.0215 & 0.0334 & 0.0362 & 0.0844 & 0.0907 & 0.0137 & 0.0121 \\
& sv\_svo & 0.2594 & 0.5456 & 0.1969 & 0.1656 & 0.0909 & 0.1461 & 0.2036 & 0.2175 & 0.3824 & 0.3892 & 0.1157 & 0.1132 \\
& sv\_svo\_svoa & 0.3005 & 0.5907 & 0.2210 & 0.1834 & 0.1067 & 0.1647 & 0.2296 & 0.2451 & 0.4243 & 0.4312 & 0.1338 & 0.1325 \\
& svoa\_conv\_msg & \textbf{0.3965} & 0.6918 & 0.2831 & 0.2319 & 0.1394 & \underline{0.2121} & 0.2950 & 0.3113 & 0.5223 & 0.5276 & 0.1812 & 0.1785 \\
& svo\_svoa\_conv\_msg & 0.3899 & \underline{0.7021} & \underline{0.2864} & \underline{0.2331} & \underline{0.1396} & 0.2109 & \underline{0.2959} & \underline{0.3117} & \underline{0.5230} & \underline{0.5284} & \underline{0.1823} & \underline{0.1792} \\
& sv\_svo\_svoa\_conv\_msg & \underline{0.3950} & \textbf{0.7092} & \textbf{0.2898} & \textbf{0.2357} & \textbf{0.1405} & \textbf{0.2133} & \textbf{0.2992} & \textbf{0.3136} & \textbf{0.5285} & \textbf{0.5337} & \textbf{0.1844} & \textbf{0.1803} \\
\bottomrule
\end{tabular}
\caption{Results for Different Embeddings and \textbf{Phi-3}}
\label{tab:phi_all_result}
\end{table*}

%% file: tables/llama.tex
\begin{table*}[htbp]
\centering
\scriptsize
\setlength{\tabcolsep}{4pt}
\setlength{\tabcolsep}{3pt} 
\renewcommand{\arraystretch}{0.95} 
\begin{tabular}{lcccccccccccccc}
\toprule
embedding & combination & acc@1 & acc@5 & p@5 & p@10 & r@5 & r@10 & ndcg@10 & ndcg@20 & mrr@10 & mrr@20 & map@10 & map@20 \\
\midrule
\multirow{6}{*}{\centering GTE}
& sv & 0.0286 & 0.0985 & 0.0233 & 0.0201 & 0.0086 & 0.0146 & 0.0231 & 0.0245 & 0.0588 & 0.0635 & 0.0091 & 0.0078 \\
& sv\_svo & 0.1965 & 0.4693 & 0.1576 & 0.1370 & 0.0731 & 0.1198 & 0.1644 & 0.1761 & 0.3131 & 0.3213 & 0.0885 & 0.0860 \\
& sv\_svo\_svoa & 0.2953 & 0.5995 & 0.2267 & 0.1896 & 0.1079 & 0.1695 & 0.2351 & 0.2478 & 0.4260 & 0.4321 & 0.1372 & 0.1339 \\
& svoa\_conv\_msg & 0.3582 & 0.6655 & 0.2699 & 0.2215 & 0.1300 & 0.2004 & 0.2783 & 0.2914 & 0.4886 & 0.4943 & 0.1691 & 0.1642 \\
& svo\_svoa\_conv\_msg & \textbf{0.3767} & \textbf{0.6735} & \textbf{0.2741} & \textbf{0.2244} & \textbf{0.1321} & \textbf{0.2021} & \textbf{0.2844} & \textbf{0.2975} & \textbf{0.5026} & \textbf{0.5086} & \textbf{0.1740} & \textbf{0.1690} \\
& sv\_svo\_svoa\_conv\_msg & \underline{0.3728} & \underline{0.6730} & \underline{0.2739} & \underline{0.2240} & \underline{0.1313} & \underline{0.2012} & \underline{0.2829} & \underline{0.2958} & \underline{0.5006} & \underline{0.5066} & \underline{0.1725} & \underline{0.1675} \\
\midrule
\multirow{6}{*}{\centering Nomic}
& sv & 0.0263 & 0.0857 & 0.0201 & 0.0177 & 0.0076 & 0.0131 & 0.0203 & 0.0219 & 0.0523 & 0.0564 & 0.0079 & 0.0069 \\
& sv\_svo & 0.2125 & 0.4579 & 0.1587 & 0.1323 & 0.0728 & 0.1147 & 0.1623 & 0.1718 & 0.3177 & 0.3254 & 0.0890 & 0.0850 \\
& sv\_svo\_svoa & 0.3193 & 0.6013 & 0.2317 & 0.1900 & 0.1097 & 0.1674 & 0.2382 & 0.2503 & 0.4403 & 0.4466 & 0.1410 & 0.1371 \\
& svoa\_conv\_msg & 0.3510 & 0.6387 & 0.2567 & 0.2093 & 0.1243 & 0.1897 & 0.2648 & 0.2776 & 0.4730 & 0.4793 & 0.1603 & 0.1558 \\
& svo\_svoa\_conv\_msg & \textbf{0.3699} & \underline{0.6527} & \underline{0.2679} & \underline{0.2165} & \underline{0.1286} & \underline{0.1938} & \underline{0.2749} & \underline{0.2887} & \underline{0.4902} & \underline{0.4968} & \underline{0.1684} & \underline{0.1638} \\
& sv\_svo\_svoa\_conv\_msg & \underline{0.3693} & \textbf{0.6630} & \textbf{0.2699} & \textbf{0.2189} & \textbf{0.1289} & \textbf{0.1949} & \textbf{0.2770} & \textbf{0.2900} & \textbf{0.4929} & \textbf{0.4991} & \textbf{0.1695} & \textbf{0.1646} \\
\midrule
\multirow{6}{*}{\centering NV-Embed}
& sv & 0.0280 & 0.0954 & 0.0228 & 0.0217 & 0.0078 & 0.0149 & 0.0240 & 0.0255 & 0.0594 & 0.0643 & 0.0093 & 0.0079 \\
& sv\_svo & 0.2025 & 0.4576 & 0.1578 & 0.1308 & 0.0711 & 0.1119 & 0.1609 & 0.1701 & 0.3139 & 0.3219 & 0.0884 & 0.0844 \\
& sv\_svo\_svoa & \textbf{0.3153} & \textbf{0.5930} & \textbf{0.2257} & \textbf{0.1843} & \textbf{0.1060} & \textbf{0.1624} & \textbf{0.2327} & \textbf{0.2425} & \textbf{0.4325} & \textbf{0.4393} & \textbf{0.1370} & \textbf{0.1319} \\
& svoa\_conv\_msg & 0.2539 & 0.5113 & 0.1856 & 0.1479 & 0.0871 & 0.1298 & 0.1880 & 0.1961 & 0.3651 & 0.3725 & 0.1075 & 0.1020 \\
& svo\_svoa\_conv\_msg & 0.2819 & 0.5567 & 0.2097 & 0.1705 & 0.0980 & 0.1494 & 0.2141 & 0.2236 & 0.3989 & 0.4061 & 0.1254 & 0.1196 \\
& sv\_svo\_svoa\_conv\_msg & \underline{0.2845} & \underline{0.5644} & \underline{0.2145} & \underline{0.1732} & \underline{0.1001} & \underline{0.1516} & \underline{0.2175} & \underline{0.2264} & \underline{0.4047} & \underline{0.4117} & \underline{0.1274} & \underline{0.1211} \\
\midrule
\multirow{6}{*}{\centering LLM2Vec}
& sv & 0.0371 & 0.1291 & 0.0309 & 0.0257 & 0.0115 & 0.0190 & 0.0299 & 0.0309 & 0.0764 & 0.0816 & 0.0119 & 0.0100 \\
& sv\_svo & 0.2274 & 0.5159 & 0.1826 & 0.1536 & 0.0841 & 0.1344 & 0.1868 & 0.1992 & 0.3509 & 0.3586 & 0.1046 & 0.1013 \\
& sv\_svo\_svoa & 0.3433 & 0.6387 & 0.2532 & 0.2047 & 0.1192 & 0.1815 & 0.2586 & 0.2716 & 0.4700 & 0.4765 & 0.1559 & 0.1516 \\
& svoa\_conv\_msg & 0.3673 & 0.6735 & 0.2694 & 0.2178 & 0.1295 & 0.1966 & 0.2765 & 0.2897 & 0.4954 & 0.5012 & 0.1670 & 0.1625 \\
& svo\_svoa\_conv\_msg & \underline{0.3885} & \underline{0.6924} & \underline{0.2836} & \underline{0.2307} & \underline{0.1350} & \underline{0.2070} & \underline{0.2921} & \underline{0.3055} & \underline{0.5169} & \underline{0.5223} & \underline{0.1790} & \underline{0.1740} \\
& sv\_svo\_svoa\_conv\_msg & \textbf{0.3902} & \textbf{0.6978} & \textbf{0.2880} & \textbf{0.2326} & \textbf{0.1365} & \textbf{0.2084} & \textbf{0.2944} & \textbf{0.3073} & \textbf{0.5188} & \textbf{0.5244} & \textbf{0.1808} & \textbf{0.1754} \\
\midrule
\multirow{6}{*}{\centering OpenAI-small}
& sv & 0.0366 & 0.1183 & 0.0275 & 0.0247 & 0.0105 & 0.0179 & 0.0283 & 0.0307 & 0.0724 & 0.0780 & 0.0110 & 0.0098 \\
& sv\_svo & 0.2179 & 0.4856 & 0.1686 & 0.1401 & 0.0784 & 0.1226 & 0.1723 & 0.1852 & 0.3290 & 0.3375 & 0.0960 & 0.0937 \\
& sv\_svo\_svoa & 0.3156 & 0.6061 & 0.2359 & 0.1943 & 0.1125 & 0.1737 & 0.2427 & 0.2559 & 0.4394 & 0.4460 & 0.1439 & 0.1409 \\
& svoa\_conv\_msg & 0.3802 & 0.6655 & 0.2703 & 0.2235 & 0.1310 & 0.2038 & 0.2827 & 0.2972 & 0.5012 & 0.5073 & 0.1727 & 0.1687 \\
& svo\_svoa\_conv\_msg & \underline{0.3867} & \underline{0.6849} & \underline{0.2820} & \underline{0.2286} & \underline{0.1351} & \underline{0.2068} & \underline{0.2905} & \underline{0.3056} & \underline{0.5118} & \underline{0.5176} & \underline{0.1796} & \underline{0.1758} \\
& sv\_svo\_svoa\_conv\_msg & \textbf{0.3905} & \textbf{0.6889} & \textbf{0.2849} & \textbf{0.2296} & \textbf{0.1364} & \textbf{0.2070} & \textbf{0.2919} & \textbf{0.3067} & \textbf{0.5166} & \textbf{0.5219} & \textbf{0.1801} & \textbf{0.1764} \\
\midrule
\multirow{6}{*}{\centering OpenAI-large}
& sv & 0.0314 & 0.1125 & 0.0265 & 0.0237 & 0.0100 & 0.0169 & 0.0268 & 0.0286 & 0.0673 & 0.0728 & 0.0104 & 0.0089 \\
& sv\_svo & 0.2174 & 0.4896 & 0.1691 & 0.1459 & 0.0792 & 0.1289 & 0.1774 & 0.1922 & 0.3340 & 0.3425 & 0.0981 & 0.0972 \\
& sv\_svo\_svoa & 0.3253 & 0.6244 & 0.2408 & 0.2005 & 0.1156 & 0.1788 & 0.2504 & 0.2640 & 0.4535 & 0.4595 & 0.1490 & 0.1459 \\
& svoa\_conv\_msg & 0.3933 & 0.7072 & 0.2922 & 0.2381 & 0.1424 & 0.2174 & 0.3016 & 0.3165 & 0.5247 & 0.5299 & 0.1863 & 0.1828 \\
& svo\_svoa\_conv\_msg & \underline{0.4050} & \textbf{0.7149} & \textbf{0.2982} & \underline{0.2418} & \textbf{0.1437} & \underline{0.2189} & \underline{0.3068} & \textbf{0.3208} & \textbf{0.5351} & \textbf{0.5400} & \underline{0.1903} & \textbf{0.1862} \\
& sv\_svo\_svoa\_conv\_msg & \textbf{0.4053} & \underline{0.7135} & \underline{0.2971} & \textbf{0.2431} & \underline{0.1427} & \textbf{0.2194} & \textbf{0.3072} & \underline{0.3207} & \underline{0.5339} & \underline{0.5388} & \textbf{0.1904} & \underline{0.1861} \\
\bottomrule
\end{tabular}
\caption{Results for Different Embeddings and \textbf{LLama-3}}
\label{tab:llama_all_result}
\end{table*}

%% file: tables/qwen.tex
\begin{table*}[htbp]
\centering
\scriptsize
\setlength{\tabcolsep}{4pt}
\setlength{\tabcolsep}{3pt} 
\renewcommand{\arraystretch}{0.95} 
\begin{tabular}{lcccccccccccccc}
\toprule
embedding & combination & acc@1 & acc@5 & p@5 & p@10 & r@5 & r@10 & ndcg@10 & ndcg@20 & mrr@10 & mrr@20 & map@10 & map@20 \\
\midrule
\multirow{6}{*}{\centering GTE}
& sv & 0.0366 & 0.1077 & 0.0267 & 0.0235 & 0.0102 & 0.0172 & 0.0272 & 0.0294 & 0.0683 & 0.0739 & 0.0110 & 0.0097 \\
& sv\_svo & 0.2139 & 0.4790 & 0.1676 & 0.1433 & 0.0774 & 0.1248 & 0.1728 & 0.1864 & 0.3271 & 0.3352 & 0.0940 & 0.0920 \\
& sv\_svo\_svoa & 0.2842 & 0.5753 & 0.2159 & 0.1793 & 0.1038 & 0.1617 & 0.2231 & 0.2376 & 0.4054 & 0.4122 & 0.1292 & 0.1273 \\
& svoa\_conv\_msg & \underline{0.3636} & 0.6638 & 0.2644 & 0.2168 & 0.1276 & 0.1969 & 0.2753 & 0.2887 & 0.4900 & 0.4960 & 0.1669 & 0.1623 \\
& svo\_svoa\_conv\_msg & \textbf{0.3693} & \textbf{0.6661} & \textbf{0.2692} & \textbf{0.2211} & \underline{0.1294} & \textbf{0.1999} & \textbf{0.2801} & \textbf{0.2932} & \textbf{0.4962} & \textbf{0.5022} & \textbf{0.1712} & \textbf{0.1659} \\
& sv\_svo\_svoa\_conv\_msg & 0.3613 & \textbf{0.6661} & \underline{0.2688} & \underline{0.2208} & \textbf{0.1296} & \underline{0.1997} & \underline{0.2786} & \underline{0.2921} & \underline{0.4916} & \underline{0.4980} & \underline{0.1696} & \underline{0.1649} \\
\midrule
\multirow{6}{*}{\centering Nomic}
& sv & 0.0331 & 0.1005 & 0.0251 & 0.0206 & 0.0099 & 0.0155 & 0.0247 & 0.0260 & 0.0635 & 0.0681 & 0.0102 & 0.0088 \\
& sv\_svo & 0.2034 & 0.4682 & 0.1640 & 0.1385 & 0.0745 & 0.1193 & 0.1673 & 0.1796 & 0.3153 & 0.3232 & 0.0918 & 0.0895 \\
& sv\_svo\_svoa & 0.2885 & 0.5813 & 0.2228 & 0.1818 & 0.1050 & 0.1608 & 0.2263 & 0.2376 & 0.4119 & 0.4180 & 0.1325 & 0.1286 \\
& svoa\_conv\_msg & 0.3393 & 0.6361 & 0.2503 & 0.2055 & 0.1218 & 0.1882 & 0.2599 & 0.2725 & 0.4654 & 0.4709 & 0.1562 & 0.1523 \\
& svo\_svoa\_conv\_msg & \underline{0.3562} & \underline{0.6558} & \underline{0.2622} & \underline{0.2122} & \underline{0.1270} & \underline{0.1917} & \underline{0.2692} & \underline{0.2819} & \underline{0.4809} & \underline{0.4872} & \underline{0.1643} & \underline{0.1594} \\
& sv\_svo\_svoa\_conv\_msg & \textbf{0.3619} & \textbf{0.6587} & \textbf{0.2645} & \textbf{0.2147} & \textbf{0.1276} & \textbf{0.1927} & \textbf{0.2720} & \textbf{0.2845} & \textbf{0.4859} & \textbf{0.4920} & \textbf{0.1662} & \textbf{0.1610} \\
\midrule
\multirow{6}{*}{\centering NV-Embed}
& sv & 0.0346 & 0.1171 & 0.0290 & 0.0252 & 0.0104 & 0.0180 & 0.0284 & 0.0298 & 0.0696 & 0.0747 & 0.0113 & 0.0097 \\
& sv\_svo & 0.2131 & 0.4784 & 0.1682 & 0.1420 & 0.0758 & 0.1216 & 0.1717 & 0.1828 & 0.3283 & 0.3361 & 0.0934 & 0.0900 \\
& sv\_svo\_svoa & \textbf{0.2862} & \textbf{0.5821} & \textbf{0.2183} & \textbf{0.1787} & \textbf{0.1035} & \textbf{0.1583} & \textbf{0.2232} & \textbf{0.2333} & \textbf{0.4099} & \textbf{0.4158} & \textbf{0.1300} & \textbf{0.1251} \\
& svoa\_conv\_msg & 0.2514 & 0.4996 & 0.1785 & 0.1450 & 0.0829 & 0.1273 & 0.1836 & 0.1908 & 0.3572 & 0.3644 & 0.1045 & 0.0988 \\
& svo\_svoa\_conv\_msg & 0.2768 & 0.5501 & 0.2075 & 0.1685 & 0.0969 & 0.1479 & 0.2112 & 0.2199 & 0.3929 & 0.3999 & 0.1233 & 0.1175 \\
& sv\_svo\_svoa\_conv\_msg & \underline{0.2825} & \underline{0.5610} & \underline{0.2125} & \underline{0.1715} & \underline{0.0992} & \underline{0.1500} & \underline{0.2154} & \underline{0.2239} & \underline{0.4008} & \underline{0.4077} & \underline{0.1263} & \underline{0.1201} \\
\midrule
\multirow{6}{*}{\centering LLM2Vec}
& sv & 0.0446 & 0.1451 & 0.0363 & 0.0311 & 0.0139 & 0.0226 & 0.0358 & 0.0383 & 0.0879 & 0.0940 & 0.0145 & 0.0127 \\
& sv\_svo & 0.2371 & 0.5304 & 0.1907 & 0.1598 & 0.0880 & 0.1390 & 0.1936 & 0.2086 & 0.3581 & 0.3663 & 0.1081 & 0.1061 \\
& sv\_svo\_svoa & 0.3142 & 0.6170 & 0.2371 & 0.1976 & 0.1131 & 0.1769 & 0.2459 & 0.2587 & 0.4427 & 0.4489 & 0.1452 & 0.1419 \\
& svoa\_conv\_msg & 0.3599 & 0.6530 & 0.2620 & 0.2139 & 0.1268 & 0.1946 & 0.2709 & 0.2843 & 0.4858 & 0.4916 & 0.1631 & 0.1586 \\
& svo\_svoa\_conv\_msg & \underline{0.3799} & \underline{0.6815} & \underline{0.2791} & \underline{0.2266} & \underline{0.1347} & \underline{0.2042} & \underline{0.2866} & \underline{0.2992} & \underline{0.5093} & \underline{0.5145} & \underline{0.1746} & \underline{0.1693} \\
& sv\_svo\_svoa\_conv\_msg & \textbf{0.3879} & \textbf{0.6864} & \textbf{0.2847} & \textbf{0.2299} & \textbf{0.1365} & \textbf{0.2065} & \textbf{0.2912} & \textbf{0.3035} & \textbf{0.5152} & \textbf{0.5203} & \textbf{0.1786} & \textbf{0.1728} \\
\midrule
\multirow{6}{*}{\centering OpenAI-small}
& sv & 0.0380 & 0.1291 & 0.0320 & 0.0281 & 0.0121 & 0.0209 & 0.0322 & 0.0344 & 0.0787 & 0.0842 & 0.0130 & 0.0113 \\
& sv\_svo & 0.2231 & 0.5021 & 0.1774 & 0.1510 & 0.0832 & 0.1326 & 0.1841 & 0.1976 & 0.3426 & 0.3507 & 0.1023 & 0.1001 \\
& sv\_svo\_svoa & 0.2942 & 0.5938 & 0.2283 & 0.1879 & 0.1093 & 0.1686 & 0.2345 & 0.2476 & 0.4220 & 0.4285 & 0.1378 & 0.1348 \\
& svoa\_conv\_msg & 0.3705 & 0.6618 & 0.2674 & 0.2194 & 0.1302 & 0.2006 & 0.2783 & 0.2931 & 0.4931 & 0.4995 & 0.1700 & 0.1663 \\
& svo\_svoa\_conv\_msg & \underline{0.3807} & \underline{0.6778} & \underline{0.2779} & \underline{0.2253} & \underline{0.1348} & \underline{0.2050} & \underline{0.2861} & \underline{0.3001} & \underline{0.5056} & \underline{0.5112} & \underline{0.1760} & \underline{0.1718} \\
& sv\_svo\_svoa\_conv\_msg & \textbf{0.3816} & \textbf{0.6824} & \textbf{0.2800} & \textbf{0.2283} & \textbf{0.1358} & \textbf{0.2069} & \textbf{0.2890} & \textbf{0.3032} & \textbf{0.5082} & \textbf{0.5136} & \textbf{0.1779} & \textbf{0.1738} \\
\midrule
\multirow{6}{*}{\centering OpenAI-large}
& sv & 0.0428 & 0.1322 & 0.0333 & 0.0293 & 0.0119 & 0.0207 & 0.0338 & 0.0356 & 0.0846 & 0.0904 & 0.0137 & 0.0116 \\
& sv\_svo & 0.2291 & 0.5124 & 0.1815 & 0.1539 & 0.0844 & 0.1356 & 0.1871 & 0.2004 & 0.3488 & 0.3558 & 0.1036 & 0.1017 \\
& sv\_svo\_svoa & 0.3088 & 0.5950 & 0.2274 & 0.1889 & 0.1089 & 0.1702 & 0.2368 & 0.2510 & 0.4304 & 0.4371 & 0.1392 & 0.1367 \\
& svoa\_conv\_msg & 0.3873 & 0.6935 & 0.2852 & 0.2340 & 0.1407 & 0.2139 & 0.2966 & 0.3124 & 0.5186 & 0.5242 & 0.1828 & 0.1798 \\
& svo\_svoa\_conv\_msg & \underline{0.3965} & \underline{0.7004} & \underline{0.2922} & \underline{0.2366} & \textbf{0.1422} & \underline{0.2145} & \underline{0.3005} & \underline{0.3154} & \underline{0.5255} & \underline{0.5312} & \underline{0.1863} & \underline{0.1822} \\
& sv\_svo\_svoa\_conv\_msg & \textbf{0.4045} & \textbf{0.7021} & \textbf{0.2926} & \textbf{0.2380} & \textbf{0.1422} & \textbf{0.2157} & \textbf{0.3026} & \textbf{0.3166} & \textbf{0.5305} & \textbf{0.5358} & \textbf{0.1874} & \textbf{0.1832} \\
\bottomrule
\end{tabular}
\caption{Results for Different Embeddings and \textbf{Qwen-2}}
\label{tab:qwen_all_result}
\end{table*}

%% file: tables/gpt.tex
\begin{table*}[htbp]
\centering
\scriptsize
\setlength{\tabcolsep}{4pt}
\setlength{\tabcolsep}{3pt} 
\renewcommand{\arraystretch}{0.95} 
\begin{tabular}{lcccccccccccccc}
\toprule
embedding & combination & acc@1 & acc@5 & p@5 & p@10 & r@5 & r@10 & ndcg@10 & ndcg@20 & mrr@10 & mrr@20 & map@10 & map@20 \\
\midrule
\multirow{6}{*}{\centering GTE}
& sv & 0.0448 & 0.1345 & 0.0351 & 0.0292 & 0.0132 & 0.0214 & 0.0342 & 0.0363 & 0.0843 & 0.0903 & 0.0146 & 0.0126 \\
& sv\_svo & 0.2211 & 0.5024 & 0.1758 & 0.1482 & 0.0815 & 0.1291 & 0.1796 & 0.1931 & 0.3392 & 0.3475 & 0.0984 & 0.0962 \\
& sv\_svo\_svoa & 0.2913 & 0.5910 & 0.2214 & 0.1826 & 0.1057 & 0.1642 & 0.2279 & 0.2424 & 0.4183 & 0.4252 & 0.1329 & 0.1302 \\
& svoa\_conv\_msg & 0.3588 & 0.6495 & 0.2633 & 0.2160 & 0.1281 & 0.1967 & 0.2735 & 0.2882 & 0.4837 & 0.4899 & 0.1664 & 0.1623 \\
& svo\_svoa\_conv\_msg & \textbf{0.3676} & \underline{0.6621} & \textbf{0.2696} & \textbf{0.2207} & \textbf{0.1307} & \textbf{0.1992} & \textbf{0.2790} & \textbf{0.2939} & \textbf{0.4925} & \textbf{0.4982} & \textbf{0.1705} & \textbf{0.1665} \\
& si\_svo\_svoa\_conv\_msg & \underline{0.3633} & \textbf{0.6641} & \underline{0.2677} & \underline{0.2199} & \underline{0.1296} & \underline{0.1980} & \underline{0.2776} & \underline{0.2930} & \underline{0.4906} & \underline{0.4964} & \underline{0.1693} & \underline{0.1657} \\
\midrule
\multirow{6}{*}{\centering Nomic}
& sv & 0.0446 & 0.1188 & 0.0326 & 0.0267 & 0.0125 & 0.0198 & 0.0320 & 0.0341 & 0.0774 & 0.0832 & 0.0142 & 0.0123 \\
& sv\_svo & 0.2336 & 0.4959 & 0.1782 & 0.1491 & 0.0826 & 0.1294 & 0.1825 & 0.1934 & 0.3454 & 0.3533 & 0.1025 & 0.0989 \\
& sv\_svo\_svoa & 0.3156 & 0.5921 & 0.2268 & 0.1869 & 0.1075 & 0.1662 & 0.2347 & 0.2463 & 0.4340 & 0.4406 & 0.1391 & 0.1349 \\
& svoa\_conv\_msg & 0.3482 & 0.6421 & 0.2515 & 0.2060 & 0.1222 & 0.1886 & 0.2618 & 0.2753 & 0.4720 & 0.4784 & 0.1578 & 0.1539 \\
& svo\_svoa\_conv\_msg & \underline{0.3622} & \underline{0.6558} & \underline{0.2651} & \underline{0.2170} & \underline{0.1284} & \underline{0.1963} & \underline{0.2751} & \underline{0.2871} & \underline{0.4895} & \underline{0.4953} & \underline{0.1682} & \underline{0.1631} \\
& si\_svo\_svoa\_conv\_msg & \textbf{0.3696} & \textbf{0.6750} & \textbf{0.2719} & \textbf{0.2195} & \textbf{0.1307} & \textbf{0.1973} & \textbf{0.2784} & \textbf{0.2899} & \textbf{0.4975} & \textbf{0.5029} & \textbf{0.1702} & \textbf{0.1649} \\
\midrule
\multirow{6}{*}{\centering NV-Embed}
& sv & 0.0448 & 0.1308 & 0.0345 & 0.0298 & 0.0128 & 0.0213 & 0.0344 & 0.0362 & 0.0834 & 0.0891 & 0.0146 & 0.0126 \\
& sv\_svo & 0.2322 & 0.4944 & 0.1798 & 0.1508 & 0.0811 & 0.1278 & 0.1835 & 0.1929 & 0.3458 & 0.3538 & 0.1023 & 0.0976 \\
& sv\_svo\_svoa & \textbf{0.3019} & \textbf{0.5898} & \textbf{0.2235} & \textbf{0.1805} & \textbf{0.1057} & \textbf{0.1604} & \textbf{0.2283} & \textbf{0.2405} & \textbf{0.4244} & \textbf{0.4309} & \textbf{0.1341} & \textbf{0.1301} \\
& svoa\_conv\_msg & 0.2471 & 0.4996 & 0.1790 & 0.1452 & 0.0834 & 0.1272 & 0.1831 & 0.1897 & 0.3551 & 0.3625 & 0.1038 & 0.0981 \\
& svo\_svoa\_conv\_msg & 0.2825 & 0.5470 & 0.2056 & 0.1696 & 0.0951 & 0.1480 & 0.2121 & 0.2212 & 0.3965 & 0.4033 & 0.1233 & 0.1177 \\
& si\_svo\_svoa\_conv\_msg & \underline{0.2896} & \underline{0.5590} & \underline{0.2095} & \underline{0.1720} & \underline{0.0971} & \underline{0.1496} & \underline{0.2153} & \underline{0.2255} & \underline{0.4035} & \underline{0.4104} & \underline{0.1251} & \underline{0.1198} \\
\midrule
\multirow{6}{*}{\centering LLM2Vec}
& sv & 0.0634 & 0.1680 & 0.0448 & 0.0381 & 0.0174 & 0.0284 & 0.0454 & 0.0484 & 0.1102 & 0.1174 & 0.0200 & 0.0175 \\
& sv\_svo & 0.2585 & 0.5490 & 0.2009 & 0.1694 & 0.0926 & 0.1464 & 0.2061 & 0.2192 & 0.3811 & 0.3891 & 0.1171 & 0.1137 \\
& sv\_svo\_svoa & 0.3259 & 0.6304 & 0.2464 & 0.2024 & 0.1178 & 0.1810 & 0.2531 & 0.2667 & 0.4530 & 0.4589 & 0.1517 & 0.1485 \\
& svoa\_conv\_msg & 0.3628 & 0.6590 & 0.2610 & 0.2137 & 0.1261 & 0.1940 & 0.2709 & 0.2850 & 0.4871 & 0.4932 & 0.1633 & 0.1592 \\
& svo\_svoa\_conv\_msg & \underline{0.3813} & \underline{0.6867} & \underline{0.2800} & \underline{0.2277} & \underline{0.1344} & \underline{0.2054} & \underline{0.2884} & \underline{0.3018} & \underline{0.5101} & \underline{0.5155} & \underline{0.1767} & \underline{0.1718} \\
& si\_svo\_svoa\_conv\_msg & \textbf{0.3916} & \textbf{0.6915} & \textbf{0.2861} & \textbf{0.2319} & \textbf{0.1366} & \textbf{0.2081} & \textbf{0.2934} & \textbf{0.3068} & \textbf{0.5184} & \textbf{0.5234} & \textbf{0.1801} & \textbf{0.1749} \\
\midrule
\multirow{6}{*}{\centering OpenAI-small}
& sv & 0.0626 & 0.1520 & 0.0404 & 0.0342 & 0.0158 & 0.0250 & 0.0413 & 0.0443 & 0.1024 & 0.1089 & 0.0183 & 0.0164 \\
& sv\_svo & 0.2445 & 0.5310 & 0.1951 & 0.1616 & 0.0898 & 0.1410 & 0.1973 & 0.2097 & 0.3633 & 0.3711 & 0.1115 & 0.1084 \\
& sv\_svo\_svoa & 0.3236 & 0.6081 & 0.2329 & 0.1921 & 0.1111 & 0.1723 & 0.2419 & 0.2558 & 0.4425 & 0.4489 & 0.1432 & 0.1405 \\
& svoa\_conv\_msg & 0.3656 & 0.6664 & 0.2695 & 0.2194 & 0.1314 & 0.2006 & 0.2785 & 0.2937 & 0.4928 & 0.4987 & 0.1707 & 0.1674 \\
& svo\_svoa\_conv\_msg & \underline{0.3830} & \underline{0.6781} & \underline{0.2804} & \underline{0.2286} & \underline{0.1355} & \underline{0.2086} & \underline{0.2897} & \underline{0.3035} & \underline{0.5088} & \underline{0.5148} & \underline{0.1792} & \underline{0.1744} \\
& si\_svo\_svoa\_conv\_msg & \textbf{0.3873} & \textbf{0.6861} & \textbf{0.2844} & \textbf{0.2306} & \textbf{0.1371} & \textbf{0.2092} & \textbf{0.2928} & \textbf{0.3065} & \textbf{0.5144} & \textbf{0.5199} & \textbf{0.1812} & \textbf{0.1766} \\
\midrule
\multirow{6}{*}{\centering OpenAI-large}
& sv & 0.0531 & 0.1531 & 0.0402 & 0.0334 & 0.0151 & 0.0247 & 0.0394 & 0.0423 & 0.0962 & 0.1028 & 0.0170 & 0.0149 \\
& sv\_svo & 0.2539 & 0.5293 & 0.1934 & 0.1612 & 0.0897 & 0.1401 & 0.1982 & 0.2112 & 0.3720 & 0.3796 & 0.1118 & 0.1089 \\
& sv\_svo\_svoa & 0.3202 & 0.6121 & 0.2374 & 0.1953 & 0.1144 & 0.1759 & 0.2455 & 0.2599 & 0.4457 & 0.4518 & 0.1456 & 0.1431 \\
& svoa\_conv\_msg & 0.4013 & 0.6992 & 0.2871 & 0.2360 & 0.1414 & 0.2167 & 0.3004 & 0.3153 & 0.5286 & 0.5343 & 0.1857 & 0.1819 \\
& svo\_svoa\_conv\_msg & \underline{0.4047} & \underline{0.7112} & \underline{0.2940} & \underline{0.2381} & \underline{0.1437} & \underline{0.2168} & \underline{0.3034} & \underline{0.3180} & \underline{0.5343} & \underline{0.5395} & \underline{0.1882} & \underline{0.1844} \\
& si\_svo\_svoa\_conv\_msg & \textbf{0.4085} & \textbf{0.7124} & \textbf{0.2955} & \textbf{0.2402} & \textbf{0.1439} & \textbf{0.2178} & \textbf{0.3056} & \textbf{0.3198} & \textbf{0.5362} & \textbf{0.5415} & \textbf{0.1898} & \textbf{0.1856} \\
\bottomrule
\end{tabular}
\caption{Results for Different Embeddings and \textbf{GPT-3.5-turbo}}
\label{tab:gpt_all_result}
\end{table*}

%% file: tables/haiku.tex
\begin{table*}[htbp]
\centering
\scriptsize
\setlength{\tabcolsep}{4pt}
\setlength{\tabcolsep}{3pt} 
\renewcommand{\arraystretch}{0.95} 
\begin{tabular}{lcccccccccccccc}
\toprule
embedding & combination & acc@1 & acc@5 & p@5 & p@10 & r@5 & r@10 & ndcg@10 & ndcg@20 & mrr@10 & mrr@20 & map@10 & map@20 \\
\midrule
\multirow{6}{*}{\centering GTE}
& sv & 0.0411 & 0.1297 & 0.0316 & 0.0279 & 0.0120 & 0.0205 & 0.0320 & 0.0340 & 0.0799 & 0.0854 & 0.0129 & 0.0113 \\
& sv\_svo & 0.2228 & 0.5013 & 0.1797 & 0.1498 & 0.0829 & 0.1307 & 0.1826 & 0.1947 & 0.3431 & 0.3507 & 0.1012 & 0.0983 \\
& sv\_svo\_svoa & 0.2916 & 0.6010 & 0.2268 & 0.1876 & 0.1078 & 0.1673 & 0.2327 & 0.2469 & 0.4242 & 0.4302 & 0.1351 & 0.1324 \\
& svoa\_conv\_msg & 0.3628 & 0.6724 & 0.2668 & 0.2172 & 0.1282 & 0.1967 & 0.2748 & 0.2891 & 0.4918 & 0.4977 & 0.1658 & 0.1615 \\
& sit\_svoa\_conv\_msg & \underline{0.3665} & \underline{0.6795} & \textbf{0.2734} & \underline{0.2219} & \textbf{0.1308} & \underline{0.1997} & \underline{0.2796} & \textbf{0.2944} & \underline{0.4969} & \underline{0.5027} & \underline{0.1702} & \textbf{0.1659} \\
& sv\_svo\_svoa\_conv\_msg & \textbf{0.3716} & \textbf{0.6801} & \underline{0.2715} & \textbf{0.2234} & \underline{0.1301} & \textbf{0.2007} & \textbf{0.2809} & \underline{0.2943} & \textbf{0.4995} & \textbf{0.5054} & \textbf{0.1707} & \textbf{0.1659} \\
\midrule
\multirow{6}{*}{\centering Nomic}
& sv & 0.0394 & 0.1163 & 0.0291 & 0.0242 & 0.0107 & 0.0180 & 0.0285 & 0.0320 & 0.0726 & 0.0788 & 0.0117 & 0.0105 \\
& sv\_svo & 0.2359 & 0.4973 & 0.1787 & 0.1492 & 0.0814 & 0.1287 & 0.1828 & 0.1927 & 0.3470 & 0.3546 & 0.1028 & 0.0986 \\
& sv\_svo\_svoa & 0.3096 & 0.6147 & 0.2348 & 0.1897 & 0.1105 & 0.1669 & 0.2382 & 0.2486 & 0.4370 & 0.4429 & 0.1410 & 0.1362 \\
& svoa\_conv\_msg & 0.3365 & 0.6398 & 0.2514 & 0.2044 & 0.1220 & 0.1861 & 0.2587 & 0.2735 & 0.4647 & 0.4707 & 0.1553 & 0.1521 \\
& sit\_svoa\_conv\_msg & \underline{0.3559} & \underline{0.6561} & \underline{0.2666} & \underline{0.2175} & \underline{0.1281} & \underline{0.1948} & \underline{0.2735} & \underline{0.2866} & \underline{0.4829} & \underline{0.4886} & \underline{0.1670} & \underline{0.1620} \\
& sv\_svo\_svoa\_conv\_msg & \textbf{0.3602} & \textbf{0.6590} & \textbf{0.2695} & \textbf{0.2193} & \textbf{0.1294} & \textbf{0.1955} & \textbf{0.2757} & \textbf{0.2882} & \textbf{0.4867} & \textbf{0.4928} & \textbf{0.1687} & \textbf{0.1630} \\
\midrule
\multirow{6}{*}{\centering NV-Embed}
& sv & 0.0408 & 0.1282 & 0.0318 & 0.0291 & 0.0115 & 0.0204 & 0.0327 & 0.0343 & 0.0805 & 0.0858 & 0.0131 & 0.0113 \\
& sv\_svo & 0.2276 & 0.5056 & 0.1826 & 0.1494 & 0.0838 & 0.1285 & 0.1831 & 0.1923 & 0.3460 & 0.3537 & 0.1024 & 0.0976 \\
& sv\_svo\_svoa & \textbf{0.3079} & \textbf{0.5930} & \textbf{0.2248} & \textbf{0.1842} & \textbf{0.1084} & \textbf{0.1644} & \textbf{0.2320} & \textbf{0.2421} & \textbf{0.4306} & \textbf{0.4367} & \textbf{0.1357} & \textbf{0.1303} \\
& svoa\_conv\_msg & 0.2539 & 0.5110 & 0.1821 & 0.1460 & 0.0853 & 0.1290 & 0.1853 & 0.1921 & 0.3621 & 0.3692 & 0.1049 & 0.0990 \\
& sit\_svoa\_conv\_msg & 0.2851 & 0.5701 & 0.2127 & 0.1701 & 0.0994 & 0.1494 & 0.2138 & 0.2245 & 0.4030 & 0.4103 & 0.1239 & 0.1189 \\
& sv\_svo\_svoa\_conv\_msg & \underline{0.2873} & \underline{0.5810} & \underline{0.2179} & \underline{0.1751} & \underline{0.1016} & \underline{0.1531} & \underline{0.2191} & \underline{0.2279} & \underline{0.4094} & \underline{0.4162} & \underline{0.1273} & \underline{0.1212} \\
\midrule
\multirow{6}{*}{\centering LLM2Vec}
& sv & 0.0548 & 0.1602 & 0.0399 & 0.0365 & 0.0149 & 0.0262 & 0.0418 & 0.0455 & 0.1015 & 0.1087 & 0.0172 & 0.0153 \\
& sv\_svo & 0.2716 & 0.5658 & 0.2125 & 0.1743 & 0.0973 & 0.1506 & 0.2143 & 0.2261 & 0.3950 & 0.4021 & 0.1232 & 0.1192 \\
& sv\_svo\_svoa & 0.3390 & 0.6412 & 0.2516 & 0.2077 & 0.1195 & 0.1842 & 0.2597 & 0.2719 & 0.4662 & 0.4721 & 0.1556 & 0.1510 \\
& svoa\_conv\_msg & 0.3556 & 0.6684 & 0.2662 & 0.2143 & 0.1283 & 0.1945 & 0.2722 & 0.2848 & 0.4867 & 0.4928 & 0.1641 & 0.1587 \\
& sit\_svoa\_conv\_msg & \underline{0.3887} & \underline{0.6801} & \underline{0.2808} & \underline{0.2302} & \underline{0.1345} & \underline{0.2063} & \underline{0.2913} & \underline{0.3043} & \underline{0.5164} & \underline{0.5224} & \underline{0.1784} & \underline{0.1727} \\
& sv\_svo\_svoa\_conv\_msg & \textbf{0.3927} & \textbf{0.6944} & \textbf{0.2875} & \textbf{0.2340} & \textbf{0.1369} & \textbf{0.2087} & \textbf{0.2958} & \textbf{0.3086} & \textbf{0.5228} & \textbf{0.5282} & \textbf{0.1815} & \textbf{0.1758} \\
\midrule
\multirow{6}{*}{\centering OpenAI-small}
& sv & 0.0506 & 0.1514 & 0.0390 & 0.0331 & 0.0144 & 0.0239 & 0.0387 & 0.0422 & 0.0952 & 0.1019 & 0.0163 & 0.0145 \\
& sv\_svo & 0.2582 & 0.5207 & 0.1926 & 0.1612 & 0.0891 & 0.1401 & 0.1991 & 0.2108 & 0.3734 & 0.3805 & 0.1132 & 0.1095 \\
& sv\_svo\_svoa & 0.3213 & 0.6233 & 0.2411 & 0.1997 & 0.1140 & 0.1777 & 0.2493 & 0.2633 & 0.4486 & 0.4549 & 0.1486 & 0.1449 \\
& svoa\_conv\_msg & 0.3730 & 0.6784 & 0.2744 & 0.2207 & 0.1329 & 0.2015 & 0.2809 & 0.2954 & 0.5017 & 0.5075 & 0.1709 & 0.1674 \\
& sit\_svoa\_conv\_msg & \underline{0.3856} & \underline{0.6901} & \underline{0.2842} & \underline{0.2300} & \underline{0.1363} & \underline{0.2079} & \underline{0.2916} & \underline{0.3052} & \underline{0.5146} & \underline{0.5202} & \underline{0.1795} & \underline{0.1752} \\
& sv\_svo\_svoa\_conv\_msg & \textbf{0.3910} & \textbf{0.6978} & \textbf{0.2888} & \textbf{0.2328} & \textbf{0.1384} & \textbf{0.2092} & \textbf{0.2952} & \textbf{0.3085} & \textbf{0.5198} & \textbf{0.5251} & \textbf{0.1824} & \textbf{0.1776} \\
\midrule
\multirow{6}{*}{\centering OpenAI-large}
& sv & 0.0494 & 0.1422 & 0.0354 & 0.0316 & 0.0133 & 0.0229 & 0.0366 & 0.0396 & 0.0914 & 0.0982 & 0.0151 & 0.0132 \\
& sv\_svo & 0.2565 & 0.5367 & 0.1942 & 0.1613 & 0.0907 & 0.1408 & 0.1991 & 0.2103 & 0.3748 & 0.3823 & 0.1128 & 0.1089 \\
& sv\_svo\_svoa & 0.3276 & 0.6210 & 0.2433 & 0.2003 & 0.1164 & 0.1793 & 0.2509 & 0.2644 & 0.4543 & 0.4603 & 0.1486 & 0.1453 \\
& svoa\_conv\_msg & 0.3882 & 0.7072 & 0.2908 & 0.2355 & 0.1419 & 0.2157 & 0.2983 & 0.3132 & 0.5226 & 0.5277 & 0.1834 & 0.1794 \\
& sit\_svoa\_conv\_msg & \underline{0.3956} & \underline{0.7121} & \underline{0.2949} & \textbf{0.2389} & \underline{0.1428} & \textbf{0.2160} & \underline{0.3019} & \underline{0.3171} & \underline{0.5280} & \underline{0.5330} & \underline{0.1863} & \underline{0.1827} \\
& sv\_svo\_svoa\_conv\_msg & \textbf{0.4056} & \textbf{0.7129} & \textbf{0.2955} & \underline{0.2385} & \textbf{0.1434} & \underline{0.2155} & \textbf{0.3028} & \textbf{0.3185} & \textbf{0.5345} & \textbf{0.5395} & \textbf{0.1870} & \textbf{0.1835} \\
\bottomrule
\end{tabular}
\caption{Results for Different Embeddings and \textbf{Haiku}}
\label{tab:haiku_all_result}
\end{table*}

%% file: tables/base.tex
\begin{table*}[htbp]
\centering
\scriptsize
\setlength{\tabcolsep}{4pt}
\renewcommand{\arraystretch}{0.95}
\begin{tabular}{lcccccccccccccc}
\toprule
embedding & combination & acc@1 & acc@5 & p@5 & p@10 & r@5 & r@10 & ndcg@10 & ndcg@20 & mrr@10 & mrr@20 & map@10 & map@20 \\
\midrule
\multirow{3}{*}{GTE} 
& msg & 0.2656 & 0.5278 & 0.1950 & 0.1601 & 0.0944 & 0.1462 & 0.2025 & 0.2151 & 0.3772 & 0.3843 & 0.1169 & 0.1140 \\
& conv & 0.2336 & 0.5016 & 0.1629 & 0.1295 & 0.0767 & 0.1158 & 0.1667 & 0.1729 & 0.3482 & 0.3558 & 0.0879 & 0.0814 \\
& conv\_msg & 0.3156 & 0.6055 & 0.2323 & 0.1889 & 0.1118 & 0.1718 & 0.2394 & 0.2520 & 0.4380 & 0.4447 & 0.1404 & 0.1355 \\
\midrule
\multirow{3}{*}{Nomic} 
& msg & 0.2294 & 0.4773 & 0.1678 & 0.1391 & 0.0835 & 0.1292 & 0.1766 & 0.1892 & 0.3358 & 0.3438 & 0.0997 & 0.0980 \\
& conv & 0.2131 & 0.4730 & 0.1529 & 0.1241 & 0.0732 & 0.1122 & 0.1582 & 0.1684 & 0.3255 & 0.3345 & 0.0832 & 0.0790 \\
& conv\_msg & 0.2708 & 0.5487 & 0.1991 & 0.1631 & 0.0982 & 0.1522 & 0.2077 & 0.2221 & 0.3897 & 0.3972 & 0.1185 & 0.1165 \\
\midrule
\multirow{3}{*}{NV-Embed} 
& msg & 0.1962 & 0.4170 & 0.1368 & 0.1097 & 0.0612 & 0.0924 & 0.1385 & 0.1391 & 0.2904 & 0.2968 & 0.0738 & 0.0663 \\
& conv & 0.0808 & 0.1839 & 0.0459 & 0.0356 & 0.0213 & 0.0309 & 0.0480 & 0.0510 & 0.1249 & 0.1317 & 0.0221 & 0.0202 \\
& conv\_msg & 0.1571 & 0.3382 & 0.1049 & 0.0830 & 0.0474 & 0.0721 & 0.1074 & 0.1111 & 0.2346 & 0.2421 & 0.0563 & 0.0515 \\
\midrule
\multirow{3}{*}{LLM2Vec} 
& msg & 0.2619 & 0.5467 & 0.1991 & 0.1613 & 0.0963 & 0.1455 & 0.2032 & 0.2154 & 0.3820 & 0.3899 & 0.1160 & 0.1123 \\
& conv & 0.1537 & 0.3179 & 0.0933 & 0.0735 & 0.0459 & 0.0690 & 0.0989 & 0.1089 & 0.2259 & 0.2359 & 0.0499 & 0.0479 \\
& conv\_msg & 0.2825 & 0.5496 & 0.2008 & 0.1633 & 0.0985 & 0.1503 & 0.2092 & 0.2220 & 0.3988 & 0.4067 & 0.1189 & 0.1152 \\
\midrule
\multirow{3}{*}{OpenAI-small} 
& msg & 0.2599 & 0.5256 & 0.1945 & 0.1601 & 0.0972 & 0.1498 & 0.2031 & 0.2197 & 0.3724 & 0.3799 & 0.1181 & 0.1183 \\
& conv & 0.2705 & 0.5470 & 0.1869 & 0.1493 & 0.0902 & 0.1359 & 0.1929 & 0.2027 & 0.3863 & 0.3942 & 0.1060 & 0.1005 \\
& conv\_msg & 0.3116 & 0.6027 & 0.2280 & 0.1887 & 0.1126 & 0.1741 & 0.2390 & 0.2543 & 0.4350 & 0.4422 & 0.1408 & 0.1382 \\
\midrule
\multirow{3}{*}{OpenAI-large} 
& msg & 0.2893 & 0.5733 & 0.2146 & 0.1797 & 0.1075 & 0.1677 & 0.2272 & 0.2444 & 0.4123 & 0.4194 & 0.1331 & 0.1332 \\
& conv & 0.3073 & 0.5901 & 0.2129 & 0.1717 & 0.1025 & 0.1551 & 0.2203 & 0.2313 & 0.4263 & 0.4343 & 0.1248 & 0.1185 \\
& conv\_msg & 0.3425 & 0.6592 & 0.2591 & 0.2103 & 0.1286 & 0.1954 & 0.2670 & 0.2839 & 0.4759 & 0.4821 & 0.1601 & 0.1578 \\
\bottomrule
\end{tabular}
\caption{Baseline for conversation data retrieval performance of each embedding model}
\label{tab:base}
\end{table*}